\documentclass{JFM-FLM_Au}
\usepackage{color}

\lefttitle{L. Martínez-Ortíz and others}
\righttitle{Journal of Fluid Mechanics}

\title{Suppressing wall modes in confined rotating turbulent convection}

\author{
Lázaro Martínez-Ortíz\aff{1}\thanks{These authors contributed equally.},
Maarten Minartz\aff{1}\footnotemark[1],
Youri H. Lemm\aff{1},
Xander M. de Wit\aff{1},
Herman J. H. Clercx\aff{1}
\and
Rudie P. J. Kunnen\aff{1}
}

\affiliation{
\aff{1}Fluids and Flows Group, Department of Applied Physics and Science Education and J. M. Burgers Center for Fluid Dynamics,
Eindhoven University of Technology, P.O. Box 513, 5600 MB Eindhoven, The Netherlands
}

\corresau{Lázaro Martínez-Ortíz, \email{l.martinez.ortiz@tue.nl}}

\begin{document}
\maketitle

\begin{abstract}
In confined turbulent rotating convection, the largest vertical velocities are found near the sidewalls in the form of wave-like structures known as wall modes. These structures persist deep into the turbulent regime, bias heat transport, and disrupt bulk flow organisation through radial jets. Controlling or suppressing wall modes is, therefore, essential for accessing bulk dynamics free from wall-induced effects. Here, we combine experiments and direct numerical simulations to investigate wall modes control in cylindrical cells equipped with ring-shaped sidewall barriers. Barriers suppress vertical-velocity maxima near the sidewall and disrupt the characteristic wave-like pattern. Simulations further show that the barriers reduce the wall-mode-induced enhancement of heat transport, shifting it towards values characteristic of laterally periodic domains. The suppression efficiency is governed by the ratio of the barrier width to the wall-mode scale and is enhanced by the addition of a second barrier. In the horizontal plane, radial jet ejections are attenuated, while the time-averaged flow reveals suppression of the boundary zonal flow (BZF), a ring-shaped region of positive azimuthal velocity near the sidewall, provided measurements are taken away from the immediate vicinity of the barriers. In this region, isotherms bend toward the poorly conducting barrier, creating a local misalignment with the isobars and inducing a baroclinic flow adjacent to the barrier faces. This effect weakens with increasing barrier conductivity or smoother geometry. These results demonstrate that sidewall barriers provide a robust route for suppressing wall modes signatures in experimental turbulent rotating convection, while locally inducing secondary baroclinic flows near the barriers. Their use enables access to extreme rotating-convection regimes with reduced sidewall influence.
\end{abstract}

\begin{keywords}
Authors should not enter keywords on the manuscript, as these must be chosen by the author during the online submission process and will then be added during the typesetting process (see \href{https://www.cambridge.org/core/journals/journal-of-fluid-mechanics/information/list-of-keywords}{Keyword PDF} for the full list).  Other classifications will be added at the same time.
\end{keywords}


\section{Introduction}
\label{sec:headings}
Turbulent buoyancy-driven flows under rotation pervade natural systems, from planetary cores \citep{Aurnou2015} to extraterrestrial gas giants \citep{Heimpel2005}. For decades, the canonical configuration for the study of this phenomenon has been the rotating Rayleigh–Bénard convection (RRBC) setup \citep{Chandrasekhar1961, rossby1969study},  where a fluid layer subjected to global rotation becomes unstable under an imposed adverse temperature gradient, maintained by heating at the lower boundary and cooling at the upper boundary. The ultimate goal is to extrapolate insights from this simplified configuration to extreme large-scale flows encountered in nature, where the dominant force balance is typically the so-called geostrophic balance between pressure gradient and Coriolis force, defining the geostrophic regime of rotating convection \citep{kunnen2021geostrophic} where a wide variety of flow morphologies emerges \citep{ecke2023turbulent, kunnen2021geostrophic}. Of particular interest is the formation of large-scale vortical condensates in specific parameter regimes \citep{favier2014inverse, aguirre2020competition, van2025bridging}, associated with an inverse energy cascade in which energy is transferred from small to large scales \citep{boffetta2012two, alexakis2018cascades}, in contrast to the forward (large-to-small-scale) cascade characteristics of three-dimensional turbulence \citep{frisch1995turbulence}. These coherent structures are central to many geophysical and astrophysical flows \citep{aurnou2015rotating, marshall1999open, gill2016atmosphere}. In numerical simulations, the formation of such condensates has been readily observed \citep{stellmach2014approaching, favier2014inverse, rubio2014upscale, Guervilly2014, kunnen2016transition, aguirre2020competition}. This is often facilitated by simplifying assumptions, including the Boussinesq approximation, negligible centrifugal effects, stress-free boundary conditions, and laterally periodic domains. Together with advances in computational power, these have enabled detailed exploration of the conditions under which condensates emerge and evolve. Such idealised conditions are unattainable in laboratory experiments, where the formation of system-scale structures remains elusive.

In practice, rigid sidewalls and no-slip boundary conditions are unavoidable, introducing additional complexities such as the formation of Ekman and Stewartson boundary layers \citep{kunnen2011role}. The presence of these boundary layers gives rise to various secondary flow phenomena that perturb the bulk dynamics, such as Ekman pumping, which has been shown to delay the formation of this large-scale structure deeper into the supercritical regime \citep{aguirre2020competition}, and the so-called wall modes, which have garnered increasing attention in recent years. The latter appear as wave-like patterns adjacent to the sidewall, characterised by alternating regions of rising warm and descending cold fluid. They exhibit unexpectedly high velocities near the vertical boundaries and display a strongly dynamical and remarkably resilient behaviour, persisting even in highly turbulent regimes and under modifications of the container geometry. As a consequence of their presence, the bulk flow is disrupted and measurements of the heat flux are biased \citep{zhang2020boundary,de2020turbulent,de2023robust,favier2020robust}.

We briefly summarise the development of wall modes, from their early discovery to recent results in strongly nonlinear regimes. While rotation acts to delay the onset of convection in unbounded configurations \citep{chandrasekhar1953instability}, the presence of lateral boundaries modify this situation; linear stability and asymptotic theories predict that wall-attached convective modes are the first instability to emerge as the system transitions to convection in rotation-dominated regimes \citep{goldstein1993convection, herrmann1993asymptotic,zhang2009onset}. These modes were first observed in earlier experimental and theoretical studies \citep{zhong1991asymmetric, ecke1992hopf, liu1997eckhaus}, which reported azimuthally traveling temperature waves near the container sidewall, emerging before the onset of convection in the bulk. However, despite decades of research, their presence in more turbulent regimes or under strong rotational constraints has remained unclear. Using direct numerical simulations, \cite{HornSchmid2017} investigated a fluid with $Pr = 6.4$ in a cylindrical cell of diameter-to-height aspect ratio $\Gamma = 2$ at moderate Rayleigh numbers, and observed retrograde wall-localised precessing wave modes, consistent with earlier findings. With increasing $Ra$, they further reported a transition towards mixed states, in which wall-localised structures coexist with bulk convection patterns.

Only recently have two independent studies reported robust wall modes persisting deep into the strongly rotating and turbulent regime. Experiments using pressurised sulphur hexafluoride ($\mathrm{SF_6}$) as the working fluid revealed a bimodal temperature distribution near the sidewall of a cylindrical container ($\Gamma = 0.5$), indicating the persistence of wall-attached structures well beyond the weakly nonlinear onset regime. Both measurements and supporting simulations confirmed the presence of an annular flow, accompanied by temperature variations forming an anticyclonic travelling wave near the radial boundary, which enhanced local heat transport there \citep{zhang2020boundary}. In parallel, a combined experimental and numerical study revealed a notable discrepancy in total heat transport between a narrow cylindrical domain ($\Gamma = 0.2$) and simulations performed in a laterally unconfined periodic domain \citep{de2020turbulent}. The highest velocities were observed near the sidewall, where the flow organised into two distinct regions; upward motion along one side of the cell and downward motion along the opposite side. This azimuthal wavenumber $m=1$ structure precessed in the direction opposite to the imposed background rotation. At the interface between the upwelling and downwelling regions, intermittent jets were emitted into the bulk. Wall mode fingerprints have also been identified in the mean azimuthal velocity fields as patterns known as the boundary zonal flow (BZF). Analysis of the time-averaged azimuthal velocity reveals two distinct regions; an anticyclonic bulk flow and a cyclonic ring adjacent to the sidewall, whose thickness depends on the rotation rate \citep{zhang2021boundary, zhang2024wall, wedi2022experimental}. The BZF emerges as a consequence of the wall modes. Although the mean azimuthal flow associated with the BZF is cyclonic, the corresponding temperature field exhibits an anticyclonically propagating travelling wave.

All of these intriguing behaviours have attracted significant attention from the community, prompting further studies to better understand these complex flow structures. Wall modes have now been observed not only in water and pressurised gases but also in liquid metals \citep{xu2025thermovelocimetric} and in confined magnetoconvection \citep{tao2026onset}, highlighting the remarkable universality of these phenomena. Beyond these striking findings, wall modes contaminate heat transport and bulk flow measurements relative to laterally unbounded systems, as they introduce sidewall-localised contributions absent in idealised homogeneous configurations. This makes their suppression desirable when an accurate characterisation of the bulk flow, free from wall influences, is required. This raises important questions: how stable are these wall modes and can they be suppressed or eliminated?

Using numerical simulations, \citet{favier2020robust} demonstrated that wall modes persist far beyond their linear onset. These nonlinear wall modes remain robust even in the presence of bulk turbulence. They first appear alone under weak thermal forcing and persist, becoming increasingly non-linear as buoyancy increases. More notably, the authors identified wall modes as topologically protected states, showing that they survive even under significant changes in the shape of the cylindrical container. A large vertical disruption was introduced by truncating the cylindrical container and inserting a vertical rectangular barrier attached to the circular sidewall and extending radially toward the centre of the cell (forming a Pac-Man-like geometry); contrary to expectations, the wall modes showed no sign of weakening, they wrapped around the obstacle. Similarly, \citet{madonia2021velocimetry}  and \citet{de2023robust} identified wall modes that endure deep into the turbulent regime, while also showing that, at sufficiently strong forcing, the bulk turbulence begins to interfere with their development. By analysing the interaction between wall modes and the bulk flow, they found that jets penetrating into the bulk can organise the interior into a large-scale multipolar vortex structure, whose morphology depends on the aspect ratio ($\Gamma$) of the container. This perspective suggests a possible route to mitigating the influence of wall modes on bulk flow by increasing $\Gamma$. In particular, the number of azimuthal modes (pairs of ascending and descending flows) increases with $\Gamma$, indicating a transition from a wall-dominated single-mode state in slender cylinders ($\Gamma \leq 0.5$) to progressively more complex multipolar states as the confinement is relaxed; for example, two and four modes have been reported for $\Gamma = 1$ and $\Gamma = 1.5$, respectively \citep{favier2020robust,de2023robust}. A similar trend has also been reported by \citet{zhang2024wall}. Although the larger number of modes tends to reduce their direct spatial interaction with the bulk, at higher turbulence levels they can still drive intermittent jet-like ejections that penetrate into the interior. Moreover, in wider containers centrifugal effects become increasingly important, further modifying the flow dynamics \citep{hu2022centrifugal}.

Focused on suppressing wall modes, \citet{terrien2023suppression} proposed the introduction of narrow horizontal barriers (fins) along the vertical sidewalls. Using direct numerical simulations in a rectangular domain with periodic boundary conditions in the $y$ direction and no-slip, impermeable boundaries in the $x$ and $z$ directions, with gravity acting in the negative $z$ direction, they examined configurations in which identical horizontal fins, invariant along the $y$ direction and of rectangular cross-section, were mounted on both vertical sidewalls. Two thermal boundary conditions were considered for the intrusions: (i) thermally insulating barriers with fixed temperatures imposed at the top and bottom boundaries, and (ii) thermally conducting barriers, for which the outer vertical wall remained insulated while heat was allowed to diffuse through the fin with the same thermal diffusivity as the fluid. Their results demonstrated that the detrimental contribution of wall modes to convective heat transport can be substantially reduced by introducing one or more thin horizontal fins along the lateral boundaries. Among the two configurations, the fixed-temperature barriers showed superior performance, requiring a smaller critical barrier width to stabilise the wall modes, although this boundary condition is also the least feasible for practical applications. In general, the barriers stabilised the temperature anomalies (wave-like structures) associated with the wall modes and suppressed intense vertical motions near the sidewalls. However, whether this strategy would remain effective in experimental realisations or in numerical simulations with realistic boundary conditions remains to be tested.

In this paper, we address this question by investigating the effectiveness of sidewall barriers in rotating Rayleigh–Bénard convection in a cylindrical container through a combination of experiments and numerical simulations that closely mimic the experimental setup. We examine the principal hallmarks of wall modes, including strong near-wall vertical velocities, enhanced heat transport associated with wall modes, radial jet ejections, and the azimuthal time-averaged velocity field (the BZF). Our results demonstrate that these features can be systematically controlled and, in some cases, strongly suppressed, depending on the control parameters, barrier properties, and their number. In addition, we characterise the formation of secondary azimuthal flows in the vicinity of the barriers, which arise from mismatches in thermal conductivity that distort isotherms, and we propose strategies to attenuate the resulting steady flow.

The paper is organised as follows. In \S\ref{setup} and \S\ref{DNS}, we describe the experimental and numerical methodologies, including the implementation of sidewall barriers proposed by \cite{terrien2023suppression}. The results are presented in \S\ref{results} and are organised according to the different characteristic features of the wall modes affected by the barriers. \S\ref{vertical} examines the influence of one and two barriers on the vertical velocity component, one of the most prominent signatures of wall modes. Closely related, \S\ref{heat} investigates the corresponding impact on heat-transfer efficiency, as wall modes are known to enhance convective heat transport near the sidewalls \citep{de2020turbulent}. \S\ref{radial} and \S\ref{azimuthal} address the effects of the barriers on the radial and azimuthal velocity components. Particular attention is paid to the attenuation of radial jet ejections that penetrate the bulk, as well as to the BZF \citep{zhang2020boundary}, a feature observed in the time-averaged azimuthal velocity field. \S\ref{baro} explores an unintended consequence of the barriers; the formation of a steady baroclinic flow in their vicinity. We investigate the origin and consequences of this flow and assess strategies for reducing its intensity through modifications of the barrier geometry and thermal conductivity. Finally, the paper concludes with a summary of the main findings in \S\ref{concl}.


\section{Experimental and numerical methods}
\label{meth}

In this paper, we compare results from laboratory experiments and numerical simulations. The specifics of these complementary approaches are described in the following.

\subsection{Experimental setup and procedure}
\label{setup}
The RRBC configuration consists of a fluid layer of height $H$, subjected to a vertical temperature difference $\Delta$, heated from below and cooled from above, while the entire system rotates at a constant angular velocity $\Omega$. The dynamics of the system are governed by three key dimensionless parameters: the Rayleigh number (${Ra}$), which quantifies the ratio of buoyancy to diffusive effects; the Prandtl number (${Pr}$), defined as the ratio of momentum diffusivity to thermal diffusivity; and the Ekman number (${Ek}$), which compares viscous to Coriolis forces:
\begin{equation}
Ra = \frac{\alpha g \Delta H^{3}}{\nu \kappa}, \quad 
Pr = \frac{\nu}{\kappa}, \quad 
Ek = \frac{\nu}{2 \Omega H^{2}}.
\end{equation}
Above, $\alpha$ is the thermal expansion coefficient, $g$ is the gravitational acceleration, $\nu$ is the kinematic viscosity, and $\kappa$ is the thermal diffusivity of the fluid. The geometry is characterised by the aspect ratio between diameter and height, $\Gamma = D/H$, as previously defined. The relative importance of inertial and Coriolis effects is quantified by the Rossby number, $Ro = Ek \sqrt{Ra/Pr}$.

\begin{figure}[ht]
\centering
\includegraphics[width=0.9\textwidth]{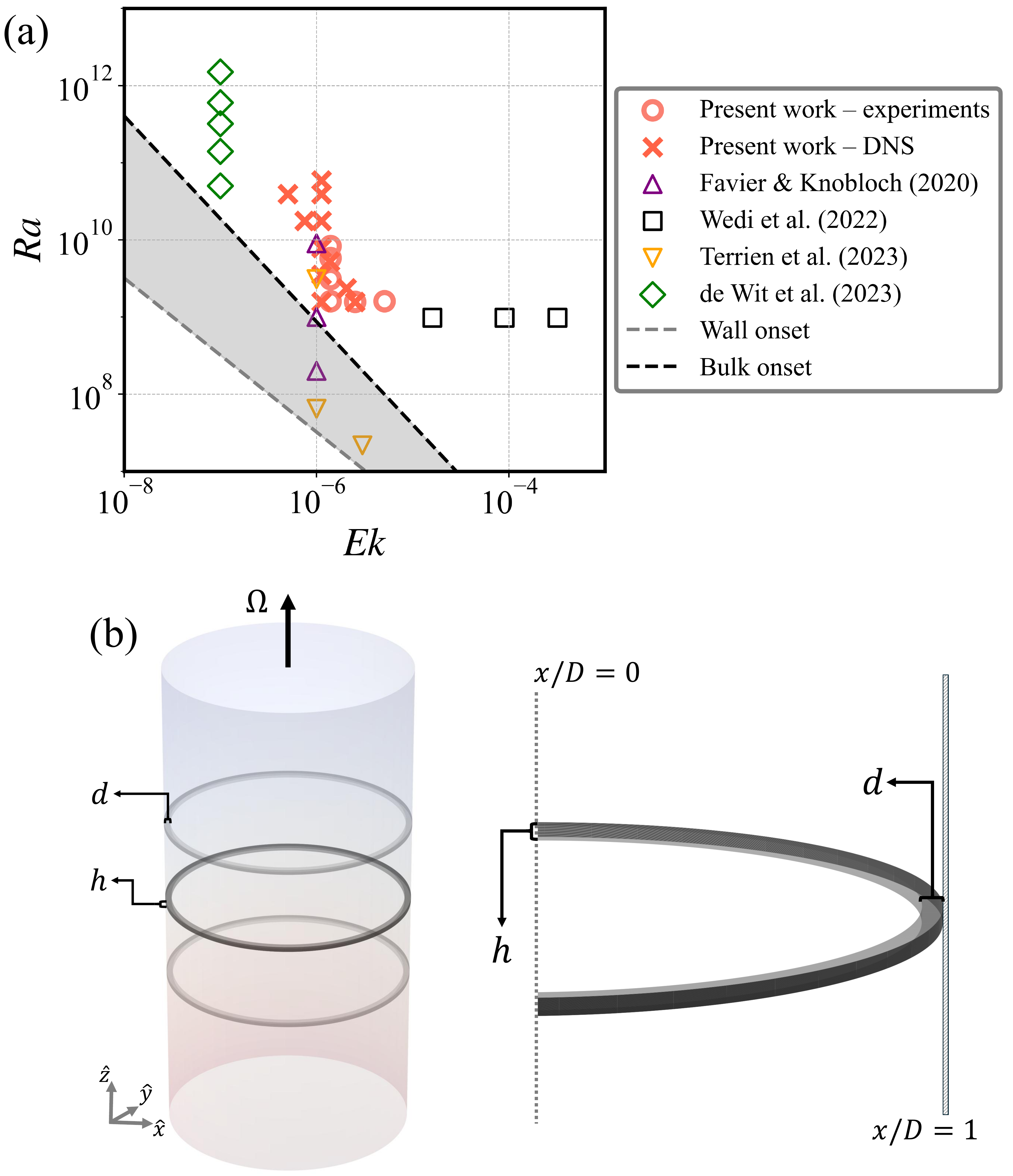}
\caption{\label{f0}(a) Parameter space $Ra$ versus $Ek$, summarizing the conducted experiments and simulations (for more details, consult Table \ref{t1} in Appendix \ref{appB}) together with results from previous studies \citep{favier2020robust,wedi2022experimental,terrien2023suppression,de2023robust}. Gray and black dashed lines show the linear stability predictions for wall modes and bulk onset, respectively. The gray-shaded region corresponds to conditions where only wall modes are expected.  
(b) Schematic of the experimental cell with sidewall barriers. A single barrier can be mounted at mid-height (dark-shaded), or alternatively, two barriers can be installed such that the distance between them and between each barrier and the top and bottom plates is $H/3$ (light-shaded). Right panel shows a zoomed-in view of the central barrier extending from the cell centre to the sidewall.
}
\end{figure}

The experiments are conducted in a cylindrical cell of diameter $D=0.2$ $\mathrm{m}$ and height $H=0.4$ $\mathrm{m}$ ($\Gamma = 0.5$), filled with water at a mean temperature $\overline{T}\approx30^\circ\mathrm{C}$. The cell consists of a copper bottom plate which is electrically heated, a transparent sapphire top plate cooled by circulating water, and a Plexiglas sidewall. The ranges of parameters explored were $1.57 \times 10^9 \leq Ra \leq 8.3 \times 10^9$, $5.0 \times 10^{-6} \leq Ek \leq 1.4 \times 10^{-6}$, with a Prandtl number of approximately $5.5$. The control parameters place the system above the onset of wall modes predicted by Herrmann \& Busse, $Ra^{\mathrm{wall}}_c \approx \pi^2(6\sqrt{3})^{1/2}\, Ek^{-1}$ \citep{herrmann1993asymptotic, zhang2017theory}, as well as above the onset of bulk convection predicted by Chandrasekhar for laterally unbounded domains, $Ra^{\mathrm{bulk}}_c \approx 3 (\pi^2/2)^{2/3} Ek^{-4/3}$ \citep{chandrasekhar1953instability, ecke2023turbulent}; see Fig. \ref{f0}(a).

The system can operate in either a smooth-wall configuration (no barriers) or with different numbers of ring-shaped barriers installed. Ring-shaped barriers, each of width $d = 10\,\mathrm{mm}$ ($d/H=0.025$, $d/D=0.05$) and height $h = 8\,\mathrm{mm}$, are mounted along the inner sidewall at various vertical positions; see Fig. \ref{f0}(b). In the single-barrier configuration, a barrier was placed at the midheight ($H/2$) of the cell (see Fig.\ref{f0}(b), unshaded). In the two barrier configuration, the barriers were symmetrically positioned, with a vertical spacing of $H/3$ between them and between the upper and lower plates (see Fig. \ref{f0}(b), shaded). Experiments were carried out using barriers made of Plexiglass (with thermal conductivity $k \approx 0.20~\mathrm{W\, m^{-1}\, K^{-1}}$ slightly lower than that of water, $k \approx 0.60~\mathrm{W\, m^{-1}\, K^{-1}}$; see \S\ref{vertical},  \S\ref{radial}, \S\ref{azimuthal}) and aluminium ($k \approx 205~\mathrm{W\, m^{-1}\, K^{-1}}$; see \S\ref{baro}). In addition, we compared barriers with straight and rounded terminations (\S\ref{baro}).

To resolve flow structures, we employ two dimensional particle image velocimetry (PIV) \citep{raffel2018piv} from both side and top views, allowing the measurement of vertical, azimuthal, and radial velocity components. A square Plexiglas jacket surrounds the cylindrical cell to minimise optical distortions caused by the curved sidewall in side-view imaging. The flow was illuminated by a $2$ mm thick light sheet generated by a $450$ nm continuous-wave diode laser, oriented vertically for side-view experiments and horizontally for top-view measurements. In the side-view configuration, the laser sheet was positioned in the vertical mid-plane ($y = D/2$), whereas in the top-view configuration it was placed at several horizontal planes at different heights illuminating $20$ $\mu$m-diameter polyamide tracer particles. Flow motion was recorded at 15 frames per second using a 5 MP monochrome camera (JAI SP-5000M-CXP2) with a resolution of $2560 \times 2048$ pixels. Time-resolved measurements were performed over more than $30$ minutes, allowing us to extract instantaneous flow structures as well as recover the time-averaged signatures of the wall modes. We performed different sets of measurements, comprising vertical and horizontal views. For the horizontal measurements, we probe the in-plane velocity field across the entire circular cross-section. Our typical velocity fields are reconstructed on $80 \times 64$ grid nodes, which corresponds to a spatial resolution of $2.5$ mm.

\subsection{Direct numerical simulations}
\label{DNS}
To access the full three-dimensional velocity and temperature fields and explore more extreme parameter regimes and barrier widths, we perform direct numerical simulations (DNS) of the incompressible Navier–Stokes equations coupled with the heat equation under the Boussinesq approximation. The equations are nondimensionalized using the temperature difference $\Delta$, the cell height $H$, and the free-fall velocity $U_{ff}=\sqrt{\alpha g \Delta H}$. No-slip boundary conditions are imposed on all boundaries, the sidewalls are thermally insulating, and the top and bottom plates are maintained at fixed temperatures. The governing equations for conservation of momentum, mass, and transport of heat read:

\begin{equation}
\frac{\partial \tilde{\mathbf{u}}}{\partial t}
+ (\tilde{\mathbf{u}} \cdot \boldsymbol{\nabla})\tilde{\mathbf{u}}
+ \frac{1}{Ek}\sqrt{\frac{Pr}{Ra}}\,\hat{\mathbf{z}}\times\tilde{\mathbf{u}}
= -\boldsymbol{\nabla} p
+ \sqrt{\frac{Pr}{Ra}} \nabla^2 \tilde{\mathbf{u}}
+ \theta \hat{\mathbf{z}}
- \chi \zeta \tilde{\mathbf{u}},
\label{e1}
\end{equation}

\begin{equation}
\boldsymbol{\nabla} \cdot \tilde{\mathbf{u}} = 0,
\label{e3}
\end{equation}

\begin{equation}
\frac{\partial \theta}{\partial t}
+ (\tilde{\mathbf{u}} \cdot \boldsymbol{\nabla})\theta
= \frac{1}{\sqrt{Pr Ra}} \nabla^2 \theta.
\label{e2}
\end{equation}

where $\tilde{\mathbf{u}}$ is the velocity field, $\theta$ the temperature, $p$ the pressure, $t$ the time, and $\hat{\mathbf{z}}$ the vertical unit vector. The ring-shaped sidewall barriers are implemented using the so called volume penalisation method \citep{kadoch2012volume, engels2015numerical}, whereby an additional term ($-\chi\zeta\mathbf{\tilde{u}}$) in Eq. (\ref{e1}) enforces vanishing velocity inside the barrier region; here $\chi=1$ inside the barrier and $\chi=0$ elsewhere. The penalisation strength is set to $\zeta=100$ for all runs, as was found adequate in our previous study using the same method \citep{martinez2025persistent}. Heat diffusion inside the barrier is treated identically to that in the fluid, using the same thermal diffusivity, consistent with Eq. (\ref{e2}). The equations are discretised on a cylindrical grid using a second-order finite-difference scheme with third-order Runge–Kutta time integration \citep{verzicco1996cylindrical,verzicco2003RBC}.

Two sets of simulations are performed. The first set, referred to as the ``constant-$Ro$ series'', spans
$1.57 \times 10^9 \leq Ra \leq 3.93 \times 10^{10}$ and
$2.52 \times 10^{-6} \leq Ek \leq 5.04 \times 10^{-7}$, while $Ro = 4.27 \times 10^{-2}$ is kept constant. In the second set, referred to as the ``constant-$Ek$ series'', simulations are conducted at fixed $Ek = 1.13 \times 10^{-6}$, with
$1.57 \times 10^9 \leq Ra \leq 5.88 \times 10^{10}$, corresponding to
$1.91 \times 10^{-2} \leq Ro \leq 1.17 \times 10^{-1}$. In all cases, the Prandtl number is fixed at $Pr = 5.5$. These parameter ranges are chosen to remain within and mimic the experimental conditions. A set of cases corresponding to a more extreme parameter regime ($Ek=10^{-7}$, $5.00 \times 10^{10} \leq Ra \leq 1.50 \times 10^{12}$, $\Gamma = 0.2$) is discussed in Appendix \ref{tro}. The specifications of all simulation runs can be found in Appendix \ref{appB}.

\section{Results}
\label{results}

In this section, we present the main findings of the study. We first examine the influence of sidewall barriers on the large vertical velocities near the sidewalls (\S\ref{vertical}) and the resulting impact on heat-transfer efficiency (\S\ref{heat}), considering different turbulent regimes, levels of rotational confinement, numbers of barriers, and barrier widths. We then investigated the impact of the barriers on the remaining velocity components, focusing on radial jet ejections into the bulk (\S\ref{radial}) and on the BZF observed in the time-averaged azimuthal velocity field (\S\ref{azimuthal}). Finally, we show that barriers also induce a steady baroclinic flow in their vicinity (\S\ref{baro}).

\subsection{Vertical velocity component}\label{vertical}

\begin{figure}[ht!]
\centering
\includegraphics[width=1\textwidth]{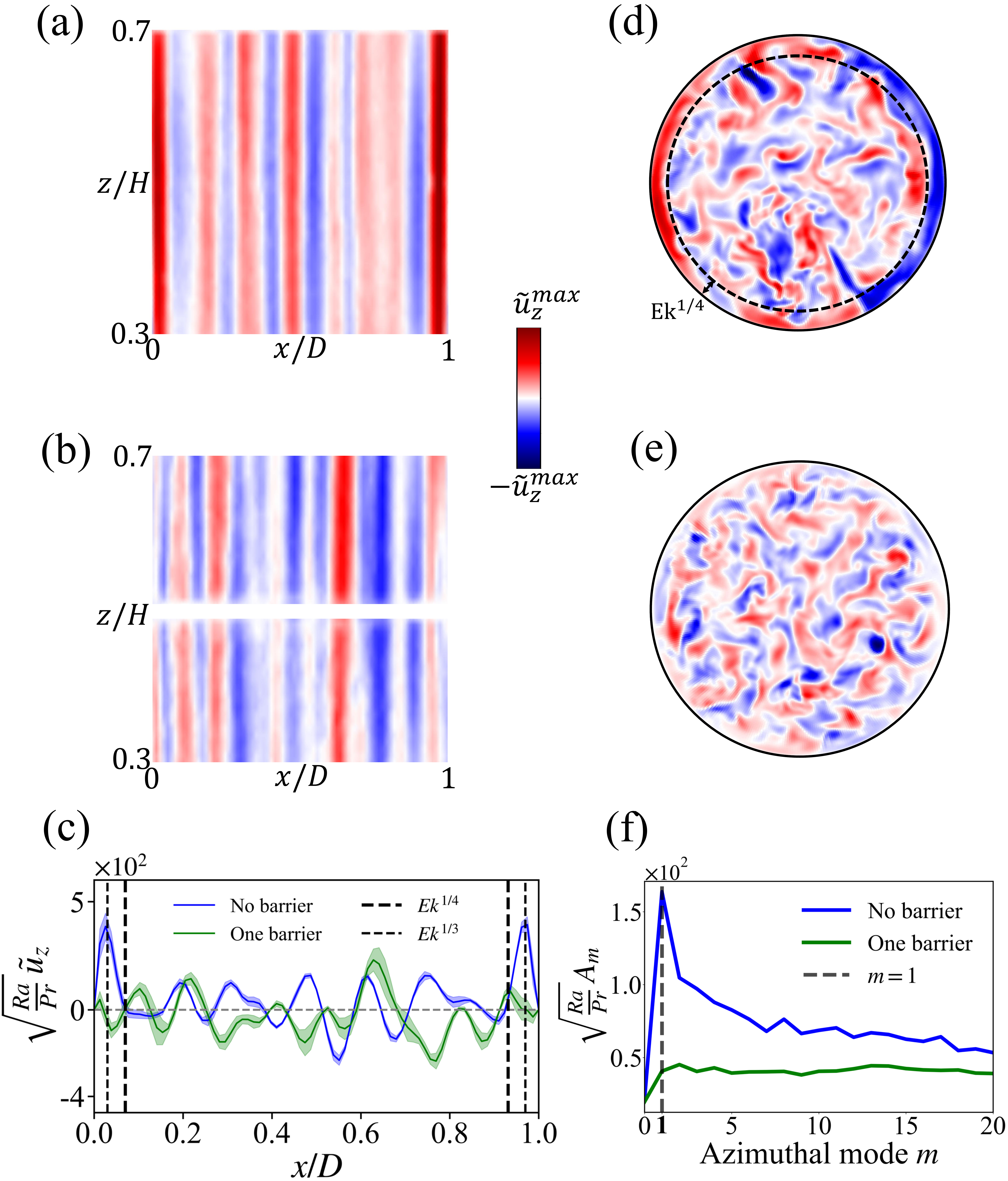}
\caption{\label{f1} (a) Vertical velocity field for the no barrier case. Strong vertical velocities are observed near the left ($x/D = 0$) and right ($x/D = 1$) walls.(b) Under the same conditions, a single barrier is placed at mid-height. Vertical velocities recover bulk values across the cross section; the barrier appears as a white band where it obstructs a small portion of the field of view.  
(c) Vertical velocity profiles along the cell diameter for both previous cases. Solid lines represent the mean profiles averaged over $0.3\le z/H\le0.7$, and the shaded regions indicate the standard deviation. These data correspond to experiments at $Ra = 1.57 \times 10^{9}$, $Ek = 1.4 \times 10^{-6}$. (d), (e) Horizontal cuts of the flow at $z=H/3$, coloured by the vertical velocity, without (d) and with (e) a barrier. The data correspond to DNS at $Ra = 7.85 \times 10^{9}$ and $Ek = 1.13 \times 10^{-6}$. (f) Time-averaged amplitudes of the azimuthal Fourier modes of the vertical velocity $\tilde{u}_z$ for the same cases, computed in the annular region $R - H\cdot Ek^{1/4} < r < R$.}
\end{figure}

For the parameters studied in this work, wall modes are expected to coexist with bulk turbulence; see Fig. \ref{f0}(a). In the vertical plane, the flow structures should exhibit strong velocities near the container sidewall, with a bulk region in the interior characterised by more moderate velocity values, all forming a vertically-aligned column-like pattern. Figure \ref{f1}(a) shows an experimental vertical velocity field averaged over a short time interval ($1$ min) for the smooth-wall case (i.e. without sidewall barriers). In this configuration, intense velocities appear near the left and right walls, reaching values exceeding twice those observed in the bulk; see Fig. \ref{f1}(c), clearly indicating the presence of wall modes. In contrast, Fig. \ref{f1}(b) shows the flow field under the same conditions but with a single barrier placed at mid-height. In this case, strong near-wall velocities are not present, and the velocity magnitude near the wall returns to values comparable to, or even lower than, those in the bulk; see Fig. \ref{f1}(c). As shown in this figure, the vertical dashed lines indicate the characteristic Stewartson-layer sandwich structure, comprising an inner layer of thickness $\delta_m/H={Ek}^{1/3}$ and a thicker outer layer of thickness $\delta/H={Ek}^{1/4}$. These length scales separate the sidewall-dominated region from the bulk flow \citep{Stewartson1957,Stewartson1966,kunnen2011role}. The former ($\delta_m$) approximately corresponds to the location of the maximum velocity, while the latter ($\delta$) coincides with the full extent of the wall modes. In what follows, we define the thickness of the wall mode region as $\delta/H = {Ek}^{1/4}$, which effectively captures their spatial extent. Our measurements do not cover the entire vertical extent of the cell; approximately $30$\% of the total cross-sectional area near the top and bottom plates remains unmeasured. However, the vertical symmetry of the flow allows us to extract conclusive results. 

Figure \ref{f1}(d) shows a horizontal cross-section at $H/3$ from a DNS snapshot, coloured by the vertical velocity. In this plane, the spatial structure of the wall mode is clearly visible. Within the near-wall region bounded by the sidewall and covering the wall-mode region $\delta$, warm ascending fluid occupies one sector of the circumference, while cold descending fluid occupies the remainder. The strongest velocity magnitudes are also confined to this near-wall region, resulting in a pronounced positive–negative velocity pattern, which is not perfectly symmetric, with one sector occupying a larger azimuthal extent than the other. The flow pattern also reveals that the spatial structure of the wall mode is dominated by an azimuthal wave-like pattern with $m=1$ as the dominant component. To quantify this observation, we perform an azimuthal Fourier decomposition of the vertical velocity field $\tilde{u}_z$ within the annular region bounded by the sidewall and the wall mode, defined as $R - H\cdot Ek^{1/4} < r < R$, and compute the amplitude $\sqrt{Ra/Pr}A_m$ of each azimuthal mode $m$. The resulting spectrum is shown in Fig. \ref{f1}(f). A pronounced peak at $m=1$ demonstrates that the flow is predominantly organised into a single azimuthal wave, corresponding to coherent ascending and descending fluid motions. Figure \ref{f1}(e) presents the corresponding horizontal cross-section at the same height after introducing the barrier. In this case, the region near the wall no longer exhibits a dominant $m=1$ structure, and the velocity magnitudes close to the wall become comparable to those of the bulk, in agreement with our experimental observations. The corresponding Fourier spectrum confirms that $m=1$ is no longer the dominant mode. These observations indicate that the barrier effectively suppresses the prominence of wall modes and disrupts their vertical organisation. In particular, it removes the wave-like structure in the vertical velocity field and prevents the concentration of the highest velocities near the sidewalls, providing promising evidence for the effectiveness of horizontal barriers in mitigating wall modes.

\begin{figure}[ht!]
\centering
\includegraphics[width=1\textwidth]{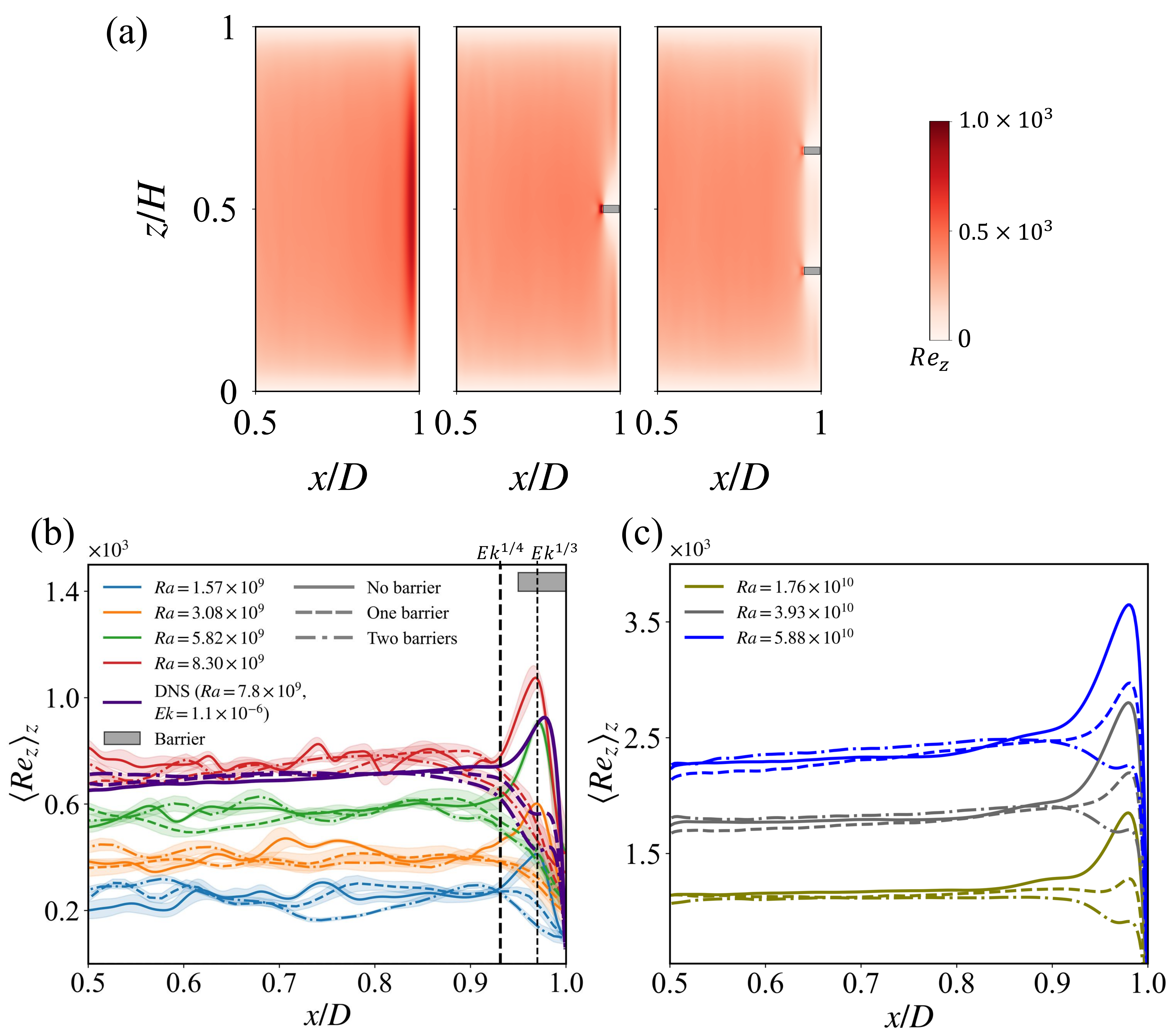}
\caption{\label{f2}(a) Azimuthally averaged DNS fields of $Re_z$ for the no-barrier, one barrier, and two barrier configurations at ${Ra}=7.8\times10^9$ and ${Ek}=1.1\times10^{-6}$. The fields are shown in cylindrical coordinates, with the symmetry axis located at $x/D=0.5$ and the sidewalls at $x/D=1$. (b) Vertically averaged $Re_z$ profiles at fixed ${Ek}=1.4\times10^{-6}$ for different values of ${Ra}$. The no barrier  cases exhibit a nearly uniform bulk region together with a pronounced near-wall peak within the $\delta$ region. The one barrier configuration strongly suppresses this near-wall peak, while the two barriers configuration leads to an even stronger suppression. The purple curves show DNS results at ${Ek}=1.1\times10^{-6}$ and ${Ra}=7.8\times10^9$, demonstrating excellent agreement with the experimental measurements. (c) Vertically averaged $Re_z$ profiles at fixed $Ek = 1.1\times10^{-6}$ and stronger thermal forcing than in (b), obtained from DNS. Although the barrier reduces the high-intensity $Re_z$ peak near the wall, a second barrier is required to further suppress the near-wall fluctuations.
} 
\end{figure}

The presence of wall modes is also reflected in the vertical velocity fluctuations, whose global structure is characterised by a relatively uniform bulk and a vertically extended region of enhanced fluctuations adjacent to the sidewall \citep{de2020turbulent}. We now examine how the sidewall barriers modify this fluctuation field and its vertically coherent structure. Figure \ref{f2}(a) shows the azimuthally averaged vertical velocity fluctuation field, $Re_z=\sqrt{Ra/Pr}~\tilde{u}^{\mathrm{rms}}_z$, over the full cell height from DNS at $Ra=7.8\times10^9$ and $Ek=1.1\times10^{-6}$. In the absence of barriers, as expected, $Re_z$ is nearly uniform throughout the bulk, while significantly enhanced fluctuations are confined to the sidewall region. Introducing a single barrier substantially reduces the magnitude of these near-wall fluctuations, while the addition of a second barrier suppresses them even further. As a result, the sidewall region becomes almost fluctuation-free compared with the no-barrier configuration. The DNS further shows that the highest vertical velocity fluctuations are now concentrated in the immediate vicinity of the barriers. A more detailed discussion of the near-barrier flow and barrier-induced effects is provided in \S\ref{baro}.

To compare the flow dynamics near the sidewall and in the bulk and quantitatively assess the effect of introducing one or two barriers, we performed long-duration experiments ($30$ min) and measured vertically averaged profiles of $Re_z$ over the accessible height range (vertical averaging denoted with $\left< \cdot \right>_z$). Figure \ref{f2}(b) shows the resulting profiles for various ${Ra}$ on a fixed ${Ek}=1.4\times10^{-6}$ for the cases without barriers, one barrier, and two barriers. In the absence of barriers, $\langle Re_z\rangle_z$ remains nearly uniform throughout the bulk, while pronounced peaks develop near the sidewall. Both bulk and sidewall values of $\langle Re_z\rangle_z$ increase with ${Ra}$, consistent with previous studies \citep{de2020turbulent}. The introduction of barriers strongly suppresses these sidewall peaks, with the two barriers configuration exhibiting the largest reduction. The effectiveness of the suppression in the strongly rotating regime can be understood from the wall-mode structure. The location of the $Re_z$ maximum from the sidewall is set by $\delta_m/H = {Ek}^{1/3}$. In our parameter range, this scale is smaller than the barrier width, $\delta_m/d < 1$. As highlighted in Fig. \ref{f2}(b), the peak of $\langle Re_z\rangle_z$ lies within the barrier region, allowing the barrier to trip the wall-modes. Introducing a single barrier is sufficient to suppress the prominent near-wall peak across all examined ${Ra}$ values, reducing $\langle Re_z\rangle_z$ within the $\delta$ region to values even below those in the bulk. In contrast, the bulk values remain largely unaffected by the presence of barriers. These results demonstrate a strong suppression of the wall-mode signature when the barrier thickness exceeds the characteristic wall-mode length scale $\delta_m$. The addition of a second barrier leads to a further reduction in $\langle Re_z\rangle_z$ within the $\delta$ region. DNS results are in good agreement with the experimental observations (purple curves in Fig. \ref{f2}(b)) and likewise exhibit a substantial reduction in the high near-wall $\langle Re_z\rangle_z$ values.

 If $Ra$ is further increased at fixed $Ek$, the DNS show that the effectiveness of the barriers decreases, consistently with \cite{terrien2023suppression}; see Fig. \ref{f2}(c) for $Ek=1.1\times10^{-6}$ and higher thermal forcing $1.76\times10^{10} \le Ra \le 5.88\times10^{10}$. Although the peak intensity of $\langle Re_z \rangle_z$ near the wall remains more than a factor of two lower than in the no barrier case for all these more extreme cases, the presence of strong fluctuations near the wall remains evident, indicating a residual signature and a re-emergence above and below the barrier of the wall modes. This highlights the need for a second barrier to further suppress wall-mode activity. The barrier thus modifies the stability of the wall modes, delaying their onset to higher supercriticality and significantly reducing their amplitude. An even more extreme set of cases, with $Ra$ up to $1.5\times10^{12}$ in a more slender cell, are presented in Appendix \ref{tro}.

\begin{figure}[ht!]
\centering
\includegraphics[width=1\textwidth]{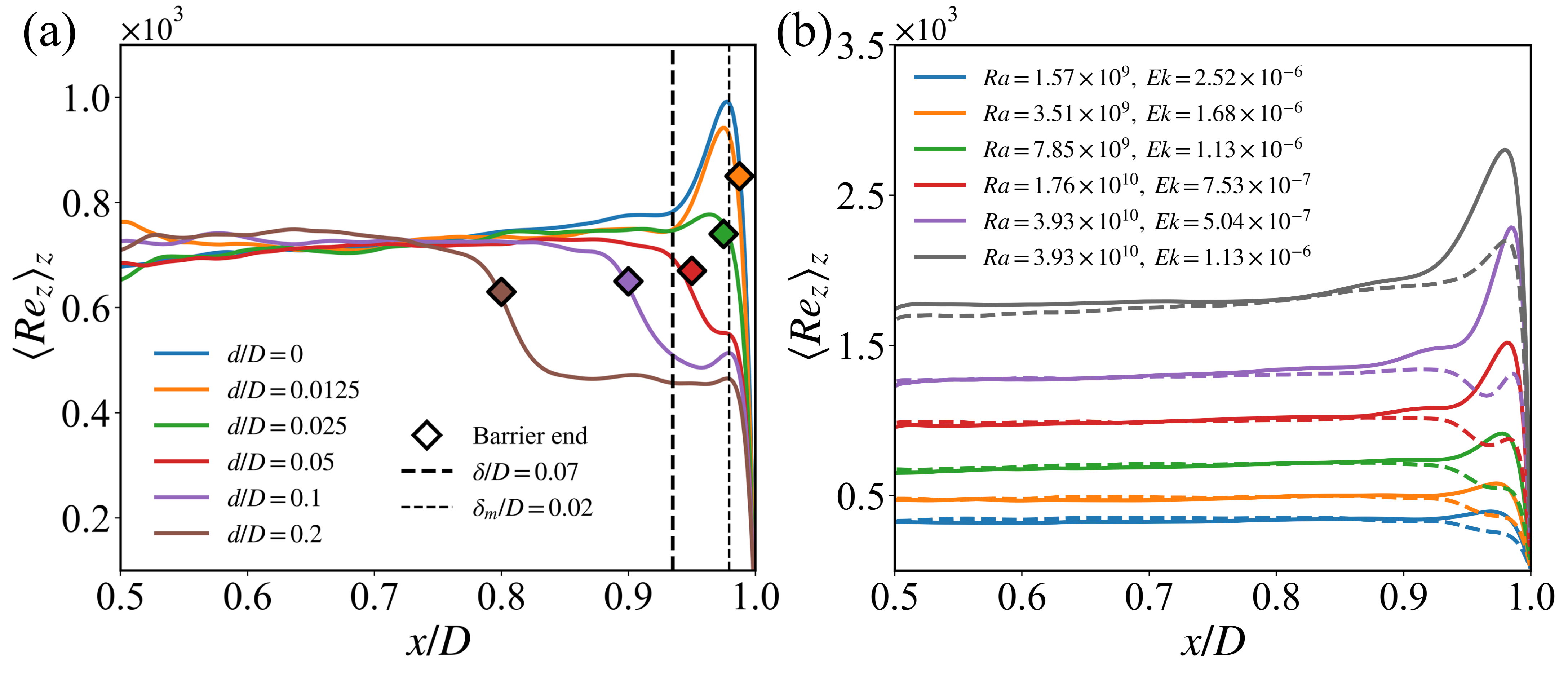}
\caption{\label{f2.1}(a) Effect of varying the barrier width at fixed ${Ra} = 7.8 \times 10^{9}$ and ${Ek} = 1.1 \times 10^{-6}$. (b) Colour profiles showing the effect of a single barrier as ${Ek}$ is decreased and ${Ra}$ is increased while maintaining ${Ro} = {Ek}\sqrt{{Ra/Pr}} = 4.27 \times 10^{-2}$. The gray profile corresponds to weaker rotational confinement at the higher Rossby number, ${Ro} = 9.55 \times 10^{-2}$. In all cases, $d/D = 0.05$ is fixed. All results are obtained from DNS.
} 
\end{figure}

The natural question that arises now is how the width of the barrier affects these results. To address this, we perform DNS in which the width of the barrier $d$ is systematically varied. Figure \ref{f2.1}(a) shows that $d$ plays a crucial role, as expected \citep{terrien2023suppression}. For a single barrier, we find that when $d$ is smaller than $\delta_m$, its influence on the wall-mode signature remains negligible. For the orange curve in Fig. \ref{f2.1}(a), corresponding to the barrier width $d/D=0.0125$, the ratio is $d/\delta_m=0.63$. In this case, no significant reduction of the near-wall $Re_z$ peak is observed. Thus, the barrier width must be comparable to or larger than the characteristic wall mode length scale, $\delta_m/H = Ek^{1/3}$, in order to reduce the maximum velocity fluctuations to levels comparable to those found in the bulk. In particular, for $d/\delta_m = 1.25$ (green curve), the $\langle Re_z\rangle_z$ peak is markedly reduced, although some residual signatures may still persist. For $d/\delta_m = 2.5$ (red curve), the $\langle Re_z\rangle_z$ peak is fully suppressed. For even larger $d$, exceeding the region $\delta/H={Ek}^{1/4}$, suppression also occurs in the bulk, which is undesirable. Introducing a second barrier accelerates these effects (not shown). These observations indicate that a single barrier strongly suppresses vertical velocities near the wall without significantly affecting the bulk when its width satisfies $\delta_m < d < \delta$.

Using DNS, we verify that this relation remains valid even at more extreme parameter values (figure \ref{f2.1}(b)). The barrier width is fixed at the experimental value ($d/D = 0.05$), while ${Ek}$ is decreased and ${Ra}$ increased, keeping constant $Ro$. As shown in figure \ref{f2.1}(b), increasing ${Ra}$ enhances the amplitude of the $\langle Re_z\rangle_z$ peaks near the wall relative to the bulk, with the maximum value reaching $1.85$ times the characteristic bulk value in the most extreme case (purple profile). However, the concurrent decrease in ${Ek}$ ensures that the characteristic wall-mode length scale, $\delta_m/H = {Ek}^{1/3}$, remains confined within the range $\delta_m < d < \delta$. As a result, a single barrier is sufficient to suppress wall modes even under these extreme conditions, with the $\langle Re_z\rangle_z$ values reduced to levels comparable to, or slightly below, those in the bulk. In contrast, the grey profile corresponds to the same ${Ra} = 3.93 \times 10^{10}$ as the purple curve, but at a higher ${Ek}$, i.e., a weaker rotational confinement. Although the overall $\langle Re_z\rangle_z$ levels are higher, the relative amplitude of the wall modes is reduced, reaching approximately $1.60$ times the bulk value. However, due to the larger horizontal extent of the wall modes at higher ${Ek}$, the introduction of a single barrier is no longer sufficient to fully suppress them, and the rms values remain approximately $1.20$ times higher in the $\delta$ region than in the bulk, suggesting the need for a wider or additional barrier.

The previous discussion provides evidence for the efficiency of barriers in controlling the most prominent feature of the wall modes, namely the high vertical velocities near the sidewalls. It also shows that predicting an exact relation for the number and characteristics of the barriers is not straightforward, as the dynamics result from a complex interplay between rotational constraints and thermal forcing. The scale of the wall modes, set by ${Ek}$, plays a dominant role; however, nonlinear effects intensified by increasing ${Ra}$ may either strengthen or cause re-emergence of high vertical-velocity fluctuations near the sidewalls. This effect is delayed by the presence of a barrier as the forcing increases; the re-emergent structures are weaker than in the configuration without a barrier at the same parameter settings, while the addition of a second barrier further enhances this weakening. Nevertheless, establishing a universal strategy to select the number and width of barriers for adequate suppression is not straightforward. Previous studies have shown that, as turbulence increases, bulk convection can eventually overtake wall modes, leading to their weakening at higher ${Ra}$ \citep{de2023robust}. These behaviours reflect the complex dynamical framework of rotating thermally driven turbulence, in which transitions between flow topologies (cells, columns, and plumes; see \cite{kunnen2021geostrophic,ecke2023turbulent}) imply that effective control strategies are likely to be regime dependent. As a result, a more extensive exploration of the parameter space and barrier characteristics is required to identify robust suppression strategies across different regimes.

\subsection{Heat transfer}\label{heat}

A direct consequence of the strong vertical velocities near the sidewalls is the enhancement of the convective heat flux in these regions \citep{zhang2020boundary,de2020turbulent,favier2020robust,zhang2021boundary,de2023robust}, which in turn biases global heat transfer in the confined configuration compared to realisations in laterally  periodic domains, where sidewalls are absent, leading to higher heat transfer efficiency in the confined case. Heat transfer is quantified by the Nusselt number, defined as $Nu= q \ H/(k\Delta)$, where $q$ is the total heat flux. It comprises both conductive and convective contributions and can be written as $q=q_{\mathrm{conv}}+q_{\mathrm{cond}}=(k/\kappa) u_z \theta-k\partial \theta/\partial z$, where the first term represents convective transport and the second term conductive transport. The heat flux is normalised by the conductive flux $q_{\mathrm{cond}}=k\Delta /H$; thus, $Nu=1$ in the absence of convection and $Nu>1$ when convection enhances heat transfer.

\begin{figure}[ht!]
\centering
\includegraphics[width=0.94\textwidth]{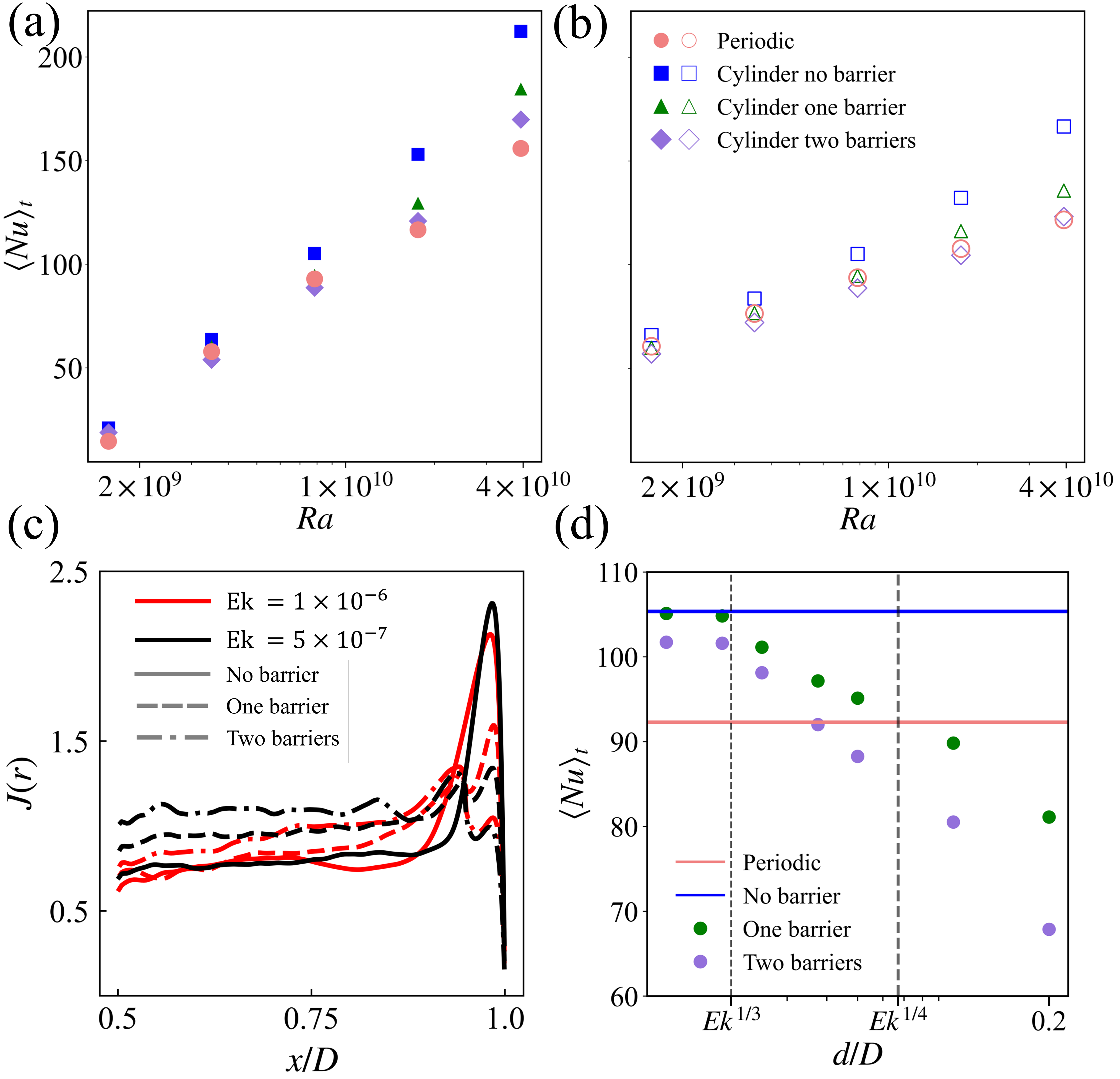}
\caption{\label{f3}(a) Dependence of the time-averaged $Nu$ on $Ra$ at fixed $Ek = 1.1 \times 10^{-6}$ from DNS.
(b) Same as in (a), but for fixed $Ro = 4.27 \times 10^{-2}$. Here, $Ra$ is varied as in (a), while $Ek$ is decreased accordingly (see Table \ref{t1}). (c) Normalized, time-, azimuthally, and vertically averaged radial profiles of heat flux, $J(r)=\langle Nu \rangle_{\phi,z,t}(r)/\langle Nu \rangle_{V,t}$, for the last set of data points in each of (a) and (b), at ${Ra}=3.93\times10^{10}$. (d) Time-averaged ${Nu}$ at fixed ${Ek} = 1.1 \times 10^{-6}$ and ${Ra} = 7.85 \times 10^9$ for various barrier widths. The cylindrical case without barrier ($d/D = 0.00$) and the periodic case are shown as horizontal lines for comparison. The barrier width increases as $d/D = 0.0125,\, 0.01875,\, 0.0250,\, 0.0375,\, 0.0500,\, 0.100,\, 0.200$.}
\end{figure}

To compare the effect of barriers on ${Nu}$ after they suppress the strong near-wall velocity signature, and to assess whether it is possible to recover values similar to the case without wall influence, we perform DNS at various ${Ra}$ while keeping ${Ek} = 1.1 \times 10^{-6}$ fixed. We consider four configurations; a horizontally periodic case, a cylindrical case without barriers, and cylindrical cases with one and two barriers along the sidewall. The results are summarised in Figure \ref{f3}(a). At low ${Ra}$, there are small differences between the periodic and cylindrical cases. However, as ${Ra}$ increases, a significant gap develops between the two, in agreement with \cite{de2020turbulent}. Barriers have a major effect on narrowing this gap. In the range $1.57 \times 10^9 \leq {Ra} \leq 7.85 \times 10^9$, a single barrier can recover values close to those of the periodic case, while adding a second barrier reduces global heat transport to values even slightly below the periodic one. As ${Ra}$ increases, it becomes more difficult to control the wall modes, as discussed previously (see Fig. \ref{f2}c). The wall modes transport heat more vigorously, making it harder for a single barrier to regulate the enhancement. However, a second barrier can reduce the gap by more than $80\%$ for ${Ra}$ up to $3.93 \times 10^{10}$, demonstrating a very strong effect on controlling this bias.

In figure \ref{f3}(b), we vary ${Ra}$ in the same range, but now at fixed ${Ro}$, such that ${Ek}$ decreases with increasing ${Ra}$. As shown previously in figure \ref{f2.1}(b), this reduces the characteristic spatial scale of the wall modes. Although their intensity increases with ${Ra}$, their tight confinement near the sidewalls allows a single barrier to effectively suppress their presence. Along this parameter cut, a single barrier is therefore more effective than in the fixed-$Ek$ case, whereas the addition of a second barrier maintains the heat transfer at a level comparable to that of the periodic case.

This is a strong result, as it is known that as the rotation becomes more dominant, the fraction of heat transported through the wall mode increases relative to the total heat transport \citep{zhang2021boundary}. This is illustrated in Fig. \ref{f3}(c), where at the same ${Ra}=3.93\times10^{10}$ the radial profiles of the local heat flux $J(r)=\langle Nu \rangle_{\phi,z,t}(r)/\langle Nu \rangle_{V,t}$ for the more rapidly rotating case exhibit a higher peak near the wall. However, at the same time, the thickness of this region decreases relative to the barrier scale, making the control more effective. In contrast, at higher ${Ek}$, the heat flux is distributed over a broader region adjacent to the wall. Therefore, after introducing a single obstruction, residual transport persists and a second barrier becomes necessary to further suppress it.

The effect of varying the barrier width at fixed ${Ek} = 1.1 \times 10^{-6}$ and ${Ra} = 7.85 \times 10^9$ is shown in figure \ref{f3}(d). When the relative width $d/D$ is smaller than the characteristic wall-mode scale, $\delta_m/H = {Ek}^{1/3}$, its influence remains weak, even in the presence of two barriers. However, as $d/D$ becomes comparable to this length scale, a single barrier is sufficient to recover $\langle{Nu}\rangle_t$ values close to those of the periodic case, while the addition of a second barrier reduces them further, even below the periodic values. For larger $d/D$, as previously noted in the vertical velocity fluctuation; see Fig. \ref{f2.1}(b), the barriers begin to interfere with bulk convection. In particular, once $d/D$ exceeds the larger characteristic scale $\delta/H={Ek}^{1/4}$, their influence extends to the bulk, leading to a significant reduction in global heat transport to values well below those of the periodic case. Such configurations are therefore unsuitable for practical applications and for comparisons with laterally unbounded systems.

Overall, these results demonstrate that sidewall barriers provide an effective means of suppressing the substantial heat-transfer bias introduced by wall modes in confined rotating convection. Similar to the behaviour of the $\langle Re_z \rangle_z$ fields, the effectiveness of the barriers is challenged as wall modes intensify with increasing thermal forcing or as their characteristic spatial scales increase, reflecting the close connection between near-wall enhancement of heat transport and strong vertical velocities. Nevertheless, barriers consistently reduce the excess heat transport associated with wall modes, bringing the global heat transfer much closer to that of the horizontally periodic cases and thereby providing a promising route for reducing wall-induced differences of global heat transport measurements in laboratory experiments.

\subsection{Radial velocity component}\label{radial}

\begin{figure}[ht]
\centering
\includegraphics[width=1\textwidth]{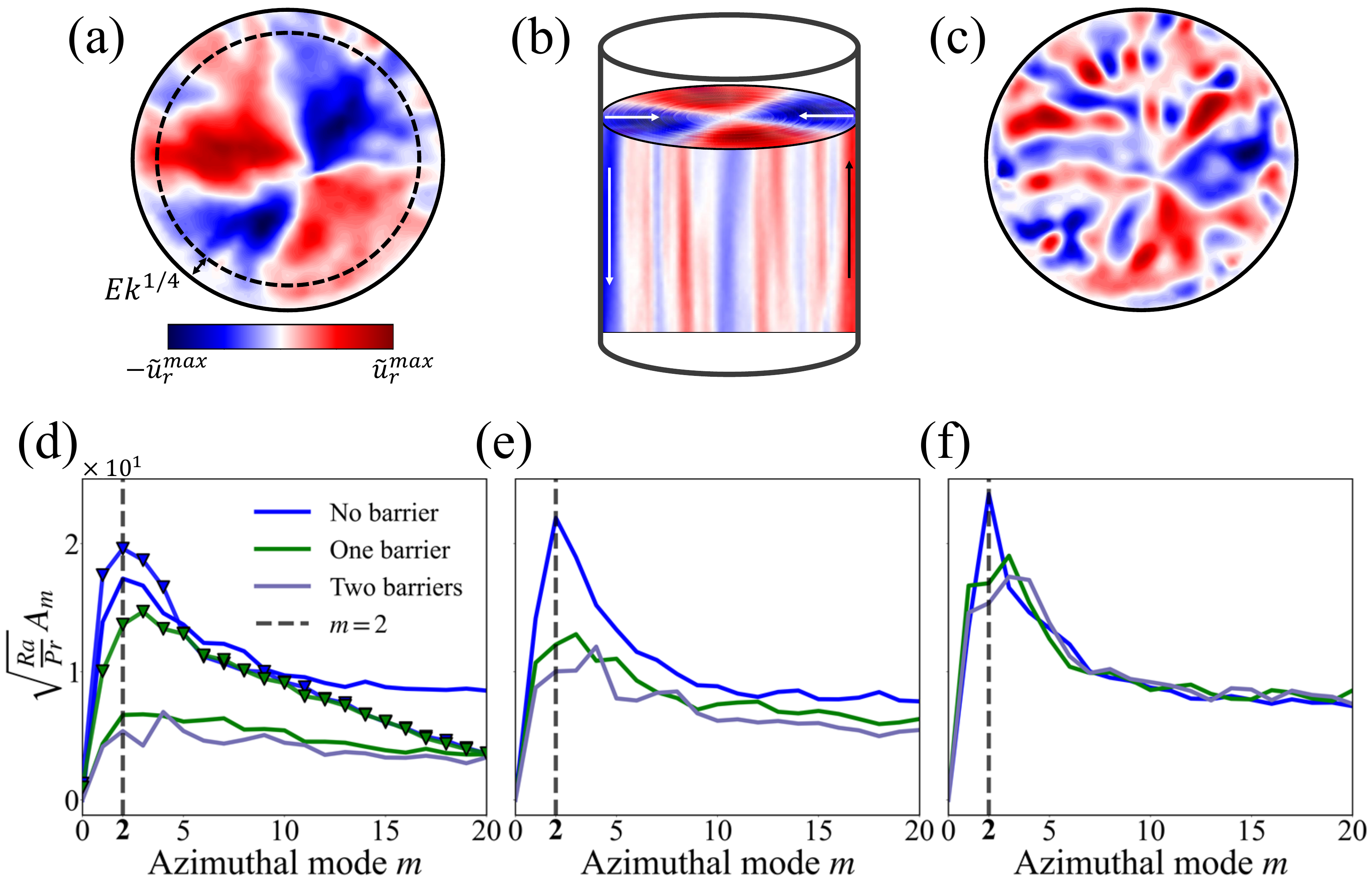}
\caption{\label{f4} (a) Instantaneous radial velocity component for ${Ra} = 1.6 \times 10^{9}$, ${Ek} = 2.5 \times 10^{-6}$, without barrier. Azimuthally, two dominant regions of opposite sign (inward and outward) are observed. (b) Full structure of the wall modes is recovered; the vertical cross-section shows maximum upward and downward velocities near the walls, while the horizontal cross-section isolates the $m=2$ Fourier component of the radial velocity, highlighting the presence of radial jets. (c) Same conditions as in (a), but with a barrier at mid-height, where smaller scales are more prominent. (d)–(f) Time-averaged Fourier spectra of the radial velocity signal for the no barrier, single- and double-barrier cases. In the no barrier case, a clear peak at $m=2$ is observed, which strengthens as the rotation rate decreases (increasing ${Ek}$). A single barrier strongly suppresses this peak at high rotation rates, although suppression is less effective at lower rotation rates, while the double-barrier configuration provides a stronger overall effect. Panels (d), (e), and (f) correspond to ${Ek} = 1.4 \times 10^{-6}$, ${Ek} = 2.5 \times 10^{-6}$, and ${Ek} = 5.0 \times 10^{-6}$, respectively, at fixed ${Ra} = 1.6 \times 10^{9}$, from experimental measurements. The lines with triangular markers in (d) correspond to DNS at ${Ek} = 1.13 \times 10^{-6}$ and ${Ra} = 7.85 \times 10^{9}$; only the no-barrier and one-barrier cases are shown for clarity.
}
\end{figure}

A particularly detrimental effect of wall modes for applications, due to their disruptive impact on flow organisation, yet also among their most striking manifestations, is the emergence of jet-like radial intrusions into the bulk as the wall modes undergo nonlinear evolution with increasing turbulence intensity \citep{de2020turbulent,favier2020robust,madonia2021velocimetry,de2023robust}. These jets emerge at the interfaces between neighbouring azimuthal mode sectors (see Figs. \ref{f1}(d) and \ref{f4}(b)) and penetrate into the cell interior, where they are deflected by the Coriolis force. This can generate multipolar large-scale flow structures, the nature of which depends on $\Gamma$. However, such dynamics impede the desired coherent organisation of the flow in highly turbulent and strongly rotating regimes.

Taking top-view measurements, Fig. \ref{f4}(a) shows, for ${Ra} = 1.6 \times 10^9$ and ${Ek} = 2.5 \times 10^{-6}$, a snapshot of the velocity field in the absence of barriers, coloured by the radial velocity component. Two pronounced inward and outward regions dominate the radial velocity field, clearly indicating the presence of intruding radial jets. A robust way to quantify this effect is, analogous to figure \ref{f1}(d)–(f), to decompose the radial velocity signal within the bulk region ($0 < r < R - H \cdot {Ek}^{1/4}$; see figure \ref{f4}(a)) into its Fourier modes. This decomposition reveals a pronounced peak in signal amplitude in the $m = 2$ mode (Fig. \ref{f4}(d)–(f)), corresponding to two regions of inward radial motion and two of outward radial motion, which are clearly identifiable in the radial velocity field. This completes the full structure of the wall modes, characterised by strong upward and downward velocities near the walls accompanied by intruding radial jets, Fig. \ref{f4}(b). To vary the ratio of the width of the barrier to the characteristic thickness of the wall mode, we vary the rotation rate (i.e., the Ekman number), while keeping ${Ra}$ fixed (see Fig. \ref{f0} (a)). This approach modifies the characteristic wall mode thickness ($\delta \sim {Ek}^{1/4}$) compared to the barrier $d$. Consequently, as $Ek$ increases, wall modes become more prominent and thicker, and the peak at $m = 2$ becomes sharper and more pronounced (see Figs. \ref{f4}(d)--(e)).

Figure \ref{f4}(c) shows snapshots of the radial velocity component at the same parameters as before, after introducing a barrier at mid-height. The flow is no longer strongly organised in a structure dominated by the $m = 2$ mode; instead, finer and more spatially fragmented scales emerge, indicating a weakening of the radial jets. In Figs. \ref{f4}(d)–(f), the effect of the barriers on the radial spectrum reveals a pronounced influence. In the faster-rotating case (lower ${Ek}$), the presence of a barrier significantly flattens the peak at $m = 2$, demonstrating a similar robust effect to that previously observed for the vertical velocity signal, where the prominent $m = 1$ structure was also suppressed in the spectrum. As ${Ek}$ increases, the wall modes become thicker, yet the barriers still reduce the peak amplitude of the $m = 2$ mode. At the highest ${Ek}$, the barriers face a more challenging regime, although they continue to provide a noticeable attenuation, with $m = 2$ no longer the dominant mode. The addition of a second barrier further aids in controlling the amplitude of the unwanted component. Using DNS, we further demonstrate that increasing ${Ra}$ (see figure \ref{f4}(d), lines with triangular markers) enhances the intensity of the component $m = 2$, consistent with the behaviours previously observed for the $\langle Re_z\rangle_z$ fields and heat transport bias (figures \ref{f2}(c) and \ref{f3}(a)). However, even in this regime, the introduction of barriers reduces the amplitude of the $m = 2$ mode such that it no longer constitutes the dominant contribution, with other components becoming comparatively more prominent.

The previously demonstrated ability to control and suppress jet ejection into the bulk is of practical relevance, as such intrusions can hinder access to and the characterisation of distinct flow regimes in the interior. Together with the disruption of the vertical organisation of the wall modes, these results demonstrate that barriers can be effectively employed to delay their nonlinear evolution and prevent their unwanted intrusion into the bulk dynamics.

\subsection{Azimuthal velocity component}\label{azimuthal}

Wall modes are also discernible in the mean fields, particularly in the azimuthal component, to which we now turn. In addition to their retrograde precession, long-time averaging of the azimuthal velocity field reveals a cyclonic, ring-shaped region adjacent to the sidewall, with a thickness set by ${Ek}$, surrounding an anticyclonic bulk region. This structure arises because cyclonic vortices preferentially cluster near the sidewall, whereas a persistent anticyclone occupies the bulk flow \citep{zhang2020boundary,zhang2021boundary,zhang2024wall,wedi2022experimental}. This pattern, referred to as the boundary zonal flow (BZF), provides an additional signature of their presence.

\begin{figure}[ht!]
\centering
\includegraphics[width=1\textwidth]{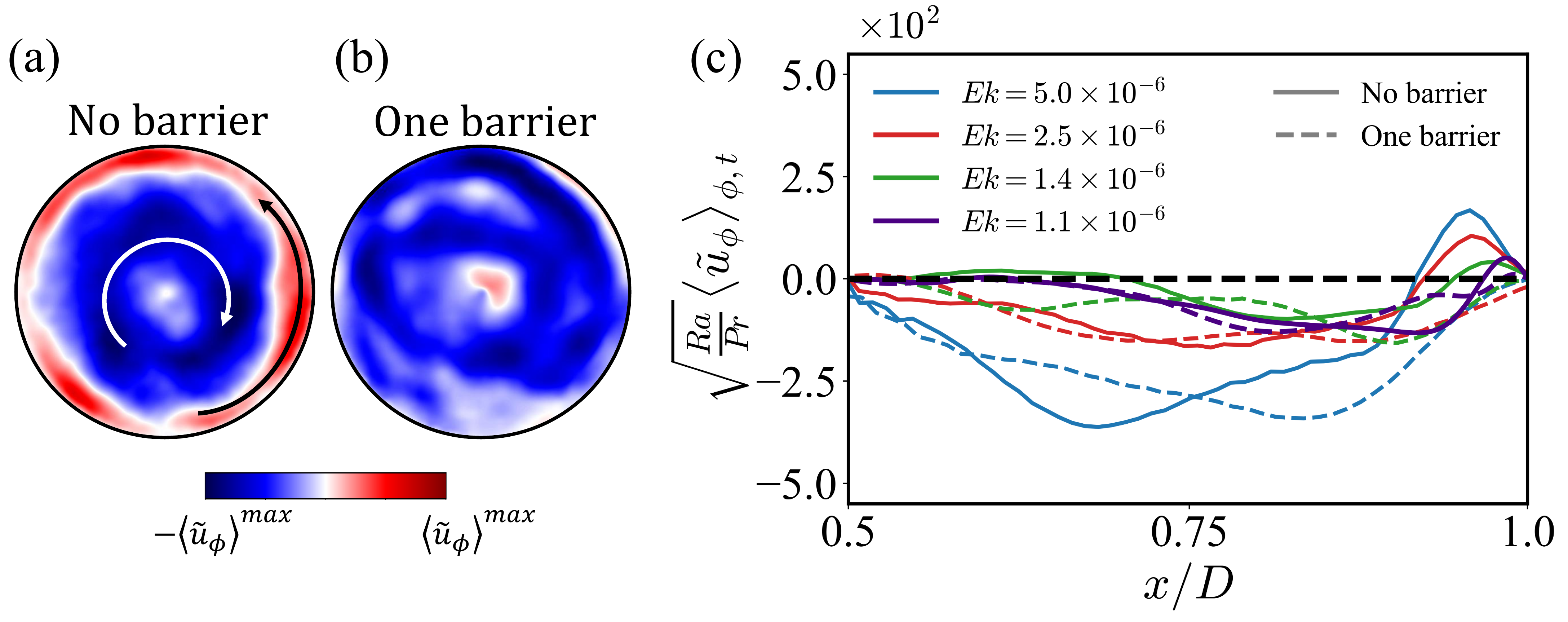}
\caption{\label{f5}
(a) Cross-section of the time-averaged azimuthal velocity field at $z = H/3$, obtained experimentally at ${Ra} = 1.57 \times 10^{9}$ and ${Ek} = 5.0 \times 10^{-6}$. An anticyclonic bulk region and a cyclonic near-wall ring corresponding to the BZF are observed. (b) Same as (a), but with one barrier. The cyclonic BZF ring is no longer present, and the mean field exhibits a fully anticyclonic cross-section. (c) Time- and azimuthally averaged $\sqrt{Ra/Pr}~\tilde{u}_\phi$ profiles. The no-barrier case shows an anticyclonic bulk region and a near-wall cyclonic region. A single barrier is enough to suppress the near-wall positive ring. Purple profiles correspond to DNS at ${Ek} = 1.1 \times 10^{-6}$ and ${Ra} = 7.85 \times 10^{9}$. The remaining profiles correspond to experiments conducted at fixed ${Ra} = 1.57 \times 10^{9}$.
}
\end{figure}

We perform long-time (30 min) averaging of the azimuthal velocity field from experiments to recover the BZF, which exhibits the expected structure: a cyclonic ring near the sidewall and an anticyclonic bulk core; see Figs. \ref{f5}(a), (c). Keeping ${Ra}$ constant while increasing ${Ek}$ leads to a thickening and increase in intensity of the BZF; see Fig. \ref{f5}(c). The introduction of a single barrier is sufficient to suppress the BZF, as shown in Fig. \ref{f5}(b) and by the dashed curves in Fig. \ref{f5}(c). The profiles in Fig. \ref{f5}(c) show that the cyclonic ring is no longer present, indicating a suppression of this wall-mode signature across all rotation rates explored. An anticyclonic bulk region remains, now extending across the entire radial domain. The addition of a second barrier further enhances this effect (not shown). In agreement with the experimental observations, the purple profiles correspond to DNS performed at lower ${Ek}$ than the experiments, resulting in a thinner BZF, while the corresponding ${Ra}$ is higher. These results further confirm that the barriers flatten the near-wall positive ring, consistent with the experimental measurements. The profiles shown here are obtained from regions away from the immediate vicinity of the barriers.

These findings provide strong evidence that the barriers also suppress the signatures of the wall modes in the mean azimuthal field, thereby eliminating the BZF. However, a relevant question that arises is whether the barriers induce secondary flows as a by-product of their presence. For the vertical and radial velocity components, the experiments show no evidence of barrier-induced flow. In the vertical component, DNS indicates that the highest vertical velocity fluctuations occur in the locality of a small region near the barrier (figure \ref{f2}(a)), a region that remains experimentally inaccessible due to barrier obstruction. In contrast, for the azimuthal component, measurements in the vicinity of the barriers reveal the development of a steady flow along the barrier faces. In the following, we examine the underlying mechanism responsible for its onset and explore possible strategies to mitigate its influence.

\subsection{Baroclinic flow induced by the barrier}\label{baro}

In \citet{terrien2023suppression}, the development of baroclinic flows around the intruding barriers was anticipated due to the lower thermal conductivity of the solid barriers compared to the surrounding fluid. These flows arise from a source term in the vorticity equation when surfaces of constant density and pressure are not aligned. Terrien and coworkers showed that for a fully insulating barrier, no static equilibrium exists because the isopycnals are misaligned with the isotherms. As a result, a steady baroclinic circulation, invariant around the barrier, spontaneously develops. To suppress this effect, the authors artificially modified the boundary conditions within the barrier: an imposed temperature barrier, with thermally insulating vertical walls and fixed temperatures at the top and bottom matching the equilibrium background, and a conducting barrier, with an insulated outer wall that allows heat to diffuse through the barrier with the same thermal diffusivity as the surrounding fluid. However, in the present experiments and in any practical implementation of this strategy, such idealised boundary conditions cannot be realised, and the resulting baroclinic flows are unavoidable.

\begin{figure}[ht!]
\centering
\includegraphics[width=1\textwidth]{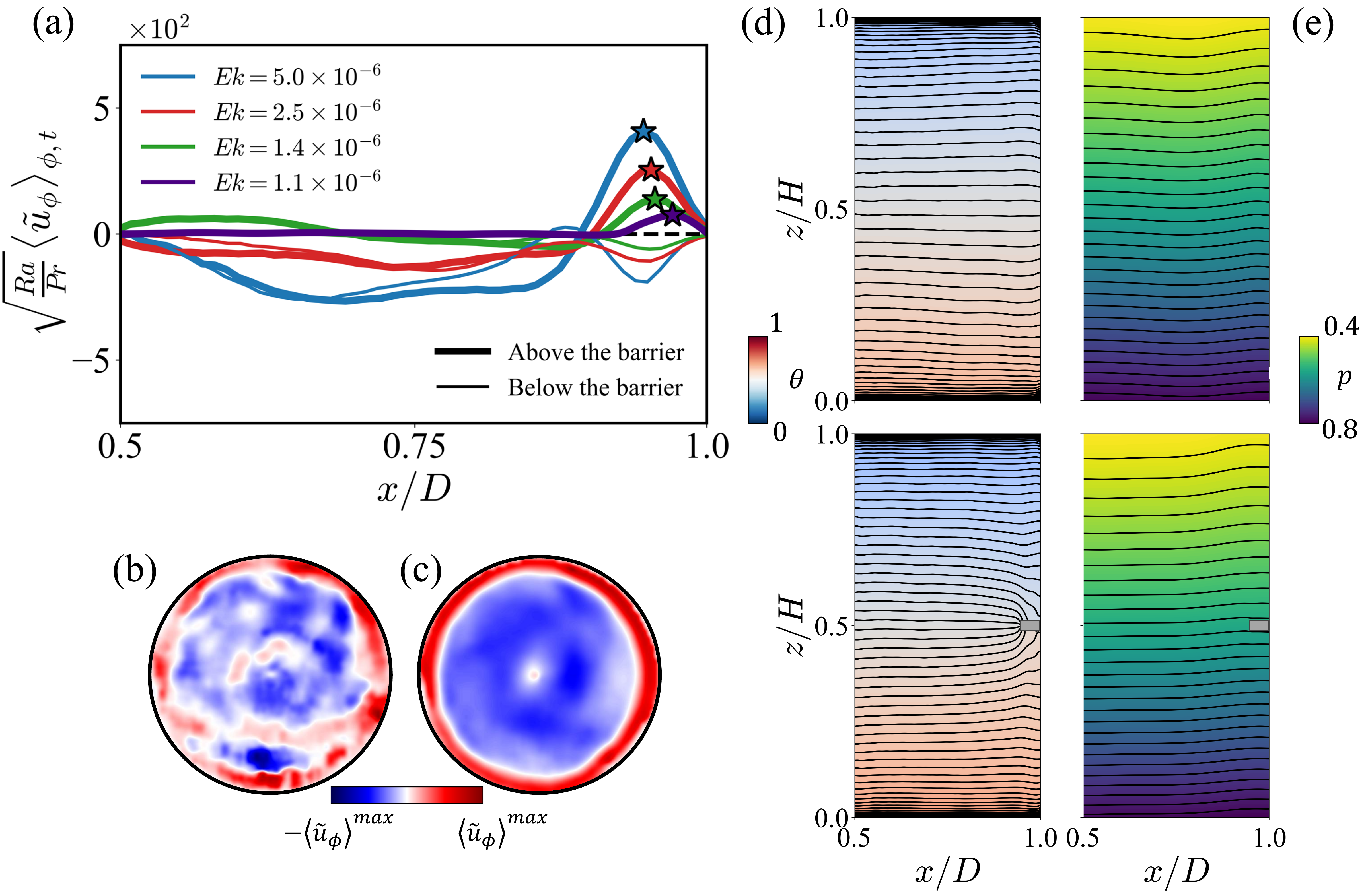}
\caption{\label{f6} (a) Time and azimuthally averaged $\sqrt{Ra/Pr}~\tilde{u}_\phi$ profiles in regions located 5 mm above and below the barrier. In both cases, an anticyclonic bulk region is observed. Above the barrier, a cyclonic near-wall region develops (stars indicate the position of the maximum in the profiles), whereas below the barrier an anticyclonic near-wall region is observed. The purple curve corresponds to the DNS; the remaining curves are from experiments, all at $Ra = 1.6 \times 10^9$. (b) Instantaneous snapshot of the azimuthal velocity field above the barrier, showing a stationary cyclonic near-wall flow at ${Ek} = 2.5\times10^{-6}$ and ${Ra} = 1.6\times10^9$. (c) Time-averaged azimuthal velocity field for the same case. (d) Temperature field from DNS, overlaid with isotherms, for the no-barrier (above) and one barrier (below) configurations. (e) Pressure field from DNS overlaid with isobars. The resulting misalignment between isotherms and isobars gives rise to the baroclinic flow.
}
\end{figure}

Figure \ref{f6}(a) shows the azimuthally averaged azimuthal velocity profiles measured in the immediate vicinity of the barrier, specifically 5 mm above and below it. Above the barrier, in its immediate vertical neighbourhood, a cyclonic steady azimuthal flow develops, whereas below the barrier an anticyclonic azimuthal flow is observed. The thickness of these layers also depends on the rotation rate, becoming thicker and of higher intensity at higher ${Ek}$ values. For the layer thickness, defined as the distance from the maximum azimuthal velocity to the sidewall, $\delta_b$, we find a scaling $\delta_b \sim {Ek}^{0.20}$ based on the four available data points, see Fig. \ref{f6}(a). Despite its similarities with the BZF, this flow is not a time-averaged effect arising from the presence of the wall modes. Rather, it is a steady flow that is already visible in the instantaneous azimuthal velocity field (Fig. \ref{f6}(b)), producing the pattern observed in the averaged field (Fig. \ref{f6}(c)).

This flow has a relatively small radial and vertical extent (see Fig. \ref{f6} in \S\ref{azimuthal} and Fig. \ref{A}(b) in Appendix \ref{appA})  and is time-independent. Moreover, it does not appear to significantly intrude into the bulk region, as indicated by the spectra shown in Figs. \ref{f4}(d)–(f), which were extracted from measurements carried out in its vicinity. However, as a warning, its intensity exceeds that of the BZF and may therefore drive unwanted dynamics. This serves as a cautionary note regarding the application of barriers, despite the positive effects discussed previously in this paper.

To understand how this stationary flow is induced by the presence of the barriers, we extracted the temperature and pressure fields from the DNS. Figure \ref{f6}(d) shows the temperature field with the overlaid isotherms. In the absence of the barrier (\ref{f6}(d) above), the isotherms remain closely spaced and approximately parallel from the bulk to the wall. However, when the barrier is present  (\ref{f6}(d) below), the isotherms bend around the barrier and intersect its surface nearly perpendicularly. In contrast, figure \ref{f6}(e) shows that the isobars remain largely unaffected by the barrier. This local misalignment between the isotherms and isobars generates the flow observed near the barrier. From the vorticity equation, one of the source terms is the baroclinic contribution $D\boldsymbol{\omega}/Dt \sim \boldsymbol{\nabla} \rho \times \boldsymbol{\nabla} p \sim \boldsymbol{\nabla} p \times \boldsymbol{\nabla} \theta$. This mechanism generates positive vorticity above the barrier and negative vorticity below it, as illustrated schematically in figure \ref{A}(a) of Appendix \ref{appA}. The resulting flow induces a symmetric inward radial velocity on both sides of the barrier, which in turn drives the steady positive azimuthal flow observed in the simulations.

So far, all experimental statistics have been well captured by numerical simulations. However, with regard to the baroclinic flow, the simulations do not reproduce the asymmetry observed in the experiments. The DNS exhibits positive azimuthal flow above and below the barrier (see Fig. \ref{f6}(a), where the thin and purple curves are on top of each other, and Fig. \ref{A}(b) of Appendix \ref{appA}), while the experiments show a clear asymmetry, with positive flow above and negative flow below. Measurements below the barrier are challenging. The Plexiglass barriers used are transparent from the top, allowing measurements beneath them; however, optical distortions, diffraction, and increased noise introduced by the barrier may partially obscure the velocity field. Nevertheless, we consistently observe this asymmetry. The differences observed between the DNS and the experiments may be influenced by additional mechanisms present in the experiments that are not captured in the numerical model, such as centrifugal effects or differences in thermal conductivity between the barrier and the fluid, which may contribute to the observed symmetry breaking. In DNS, centrifugal effects are neglected and the barrier is assumed to have the same thermal conductivity as the fluid. Incorporating these second-order effects shows that the symmetry breaking between the flows above and below the barrier can be recovered, as discussed in Appendix \ref{appA}. In general, the observed asymmetry may depend sensitively on subtle effects that are typically neglected in numerical simulations due to their expected second-order nature and the simplified formulation of the problem.

\begin{figure}[ht!]
\centering
\includegraphics[width=0.95\textwidth]{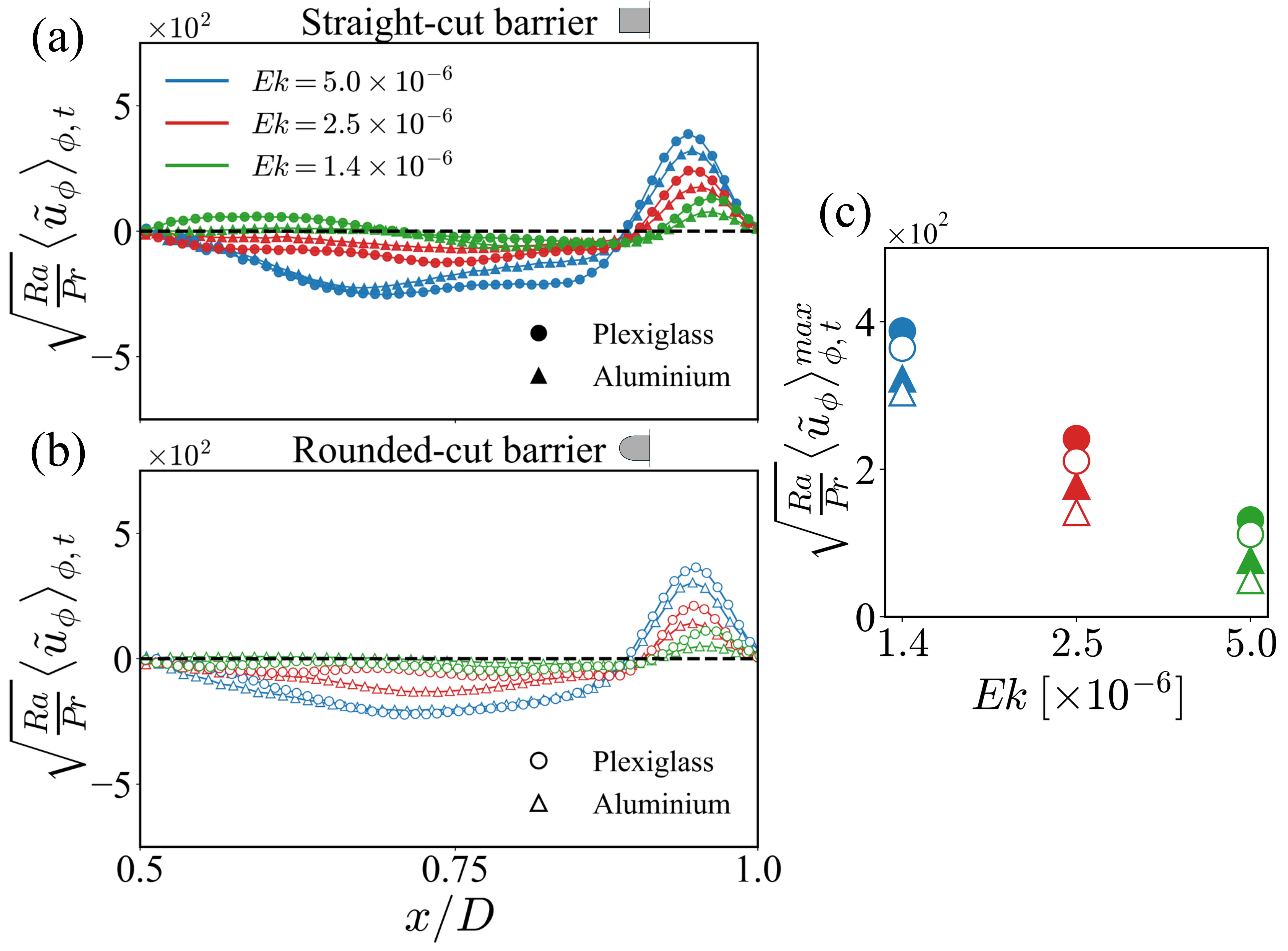}
\caption{\label{f7} 
(a) Straight-cut plexiglass and aluminium barriers are employed. Owing to the higher thermal conductivity of aluminium, the baroclinic flow intensity is reduced compared to geometrically identical plexiglass barrier.
(b) Barriers with a smoother, rounded finish are also effective; the combination of rounded geometry and aluminium material provides the strongest control, reducing the intensity of the baroclinic flow. (c) Maximum values of the profiles for each configuration at different ${Ek}$.}
\end{figure}

Although the detailed structure of this baroclinic flow warrants further investigation, its presence is expected and confirmed in the present study. We therefore also address the task of reducing its prominence. One possible solution is to minimise the abrupt bending of the isotherms near the barrier. This can be achieved by increasing the thermal conductivity of the barrier material, as the lower conductivity of the solid barrier compared to the surrounding fluid induces the observed bending of the isotherms; a higher-conductivity barrier would mitigate this effect and reduce the resulting baroclinic flow. Another approach is to prevent the isotherms from abruptly and perpendicularly intersecting the barrier surface. In the present design, the ring barriers have sharp, straight-cut terminations; smoothing these edges to produce more rounded barriers may reduce this effect by modifying the local orientation of the isotherms and avoiding their sudden perpendicular intrusion. Both strategies are tested experimentally.

Figure \ref{f7}(a) shows the first approach, in which the barrier is made of aluminium, whose thermal conductivity is approximately $10^{3}$ times higher than that of plexiglass. This modification systematically reduces the magnitude of the baroclinic flow by at least $10\%$ across all values of ${Ek}$. The second strategy, employing a rounded-cut barrier, is presented in Fig.~\ref{f7}(b) and also yields very promising results, further reducing the observed flow. The same figure demonstrates that the combination of both strategies in a rounded aluminium barrier provides the most effective configuration, with the intensity reduced by more than $15\%$ in all cases compared to the straight-cut plexiglass barriers; see Fig. \ref{f7}(c). These results indicate that both strategies reduce the strength of the baroclinic flow by limiting the strong bending of the isotherms near the barriers. High-conductivity materials appear to be the more effective strategy, while smoother barrier terminations have a more subtle effect. Further insight into how this steady flow affects bulk measurements, and how it may be further suppressed, can be developed on the basis of these initial observations.

\section{Conclusion}\label{concl}
In this study, we present an experimental evaluation of the effectiveness of sidewall barriers in suppressing wall modes in confined RRBC, supported by numerical simulations using realistic boundary conditions. We assess the efficiency of these barriers in terms of vertical, radial, and azimuthal velocity components, as well as the thermal response of the system.

In the vertical plane, vertical-velocity measurements, consistent with the DNS, show that a single barrier reduces the high near-wall velocities to values similar to or below those observed in the bulk. This is together with a disruption of the wave-like structure of the wall modes and a suppression of their high near-wall vertical-velocity fluctuations ($Re_z$ peaks) across a range of ${Ra}$ and ${Ek}$. The efficiency of these effects depends sensitively on the relative scale of the wall modes compared to the barrier width ($d)$ and on the thermal forcing, which may lead to delayed and weaker re-emergence or persistence of wall modes. By increasing the rotational confinement (i.e. decreasing $Ek$) as $Ra$ is increased, and/or by employing multiple or wider barriers, it is possible to delay the re-emergence of wall modes to deeper into the supercritical regime, while ensuring that bulk convection remains unaffected provided that ${Ek}^{1/3} \lesssim d/H \lesssim {Ek}^{1/4}$.

 DNS further shows that controlling the near-wall high velocities reduces the heat-transport bias associated with enhanced near-wall convection, recovering values comparable to those obtained under horizontally-periodic realisation. Varying the width of the barrier $d$ reveals that the wall modes effects persist when $d$ is smaller than their characteristic scale, but are progressively controlled as $d/H \gtrsim {Ek}^{1/3}$. A second barrier accelerates this suppression, enabling recovery at a smaller barrier thickness. For $d/H > Ek^{1/4}$, global heat transport falls below that obtained under periodic boundary conditions, as barriers begin to affect convection in the bulk and suppress vertical motion in this region, which is undesirable for applications. Increasing ${Ra}$ hinders suppression; however, if accompanied by a decrease in ${Ek}$, which reduces the horizontal scale of the wall-mode relative to the width of the barrier, suppression of the enhancement of the heat-transfer induced by the wall-mode remains effective even under strong thermal forcing.

As the wall modes become increasingly nonlinear, they contaminate the bulk by ejecting jets into the interior. To quantify this effect, we analyse the radial velocity from top-view measurements and examine its Fourier spectrum, which shows a dominant peak in mode $m=2$ (two inflow and two outflow regions). The introduction of a barrier strongly suppresses this peak in the more rapidly rotating cases. As expected, for thicker wall modes, the barrier becomes less effective; nevertheless, it consistently eliminates $m=2$ as the most prominent mode, while two barriers provide even stronger suppression. From the mean azimuthal velocity field, we examine another fingerprint of wall modes, the BZF, a cyclonic ring-shaped region near the wall. This structure disappears in all cases with a single barrier, regardless of wall-mode thickness, with measurements taken away from the vicinity of the barriers.

Despite their effectiveness in controlling wall-mode signatures, the barriers introduce a secondary effect. Experimental measurements reveal stationary cyclonic and anticyclonic azimuthal flows in the vicinity of the barrier, forming above and below it, respectively. By examining DNS isotherms, we find that the lower thermal transport in the barrier compared with the convecting fluid leads to local bending of the isotherms and their nearly perpendicular intersection with the barrier surfaces, while the pressure field remains essentially unaffected. This mismatch generates a baroclinic flow that manifests itself as the observed stationary azimuthal motion. In addition, we explore strategies to reduce the intensity of this flow. Two main mechanisms contribute to its formation; strong bending of isotherms near the low-conductivity barrier and their near-perpendicular intersection with its planar faces. To mitigate this, we implement aluminium barriers to increase thermal conductivity. Measurements show a clear reduction in baroclinic flow intensity of approximately $10\%$. We also test barriers with rounded edges. Combining both strategies leads to the strongest suppression of this secondary baroclinic flow (approximately $15\%$).

Overall, these results demonstrate the efficiency of sidewall barriers in suppressing wall-mode effects across a wide range of conditions, highlighting their potential as a practical tool for rapidly rotating convection experiments and simulations, where isolation of wall-mode dynamics is required. Barrier widths that remain large enough to confine the wall-mode horizontal scale while avoiding interference with bulk convection are preferable. For cases with very strong thermal forcing, the use of multiple barriers or stronger rotational confinement may be required to maintain effective suppression of wall modes. The optimal choice of barrier number and width is expected to depend strongly on the specific region of the RRBC landscape in which the system operates. As a counterpart to the presence of the barriers, a stationary azimuthal flow develops in their vicinity. The intensity of this baroclinic flow can be mitigated by employing high-thermal-conductivity materials and smooth barrier terminations.

We plan to implement these design principles in future large-scale experiments and DNS under more extreme conditions, with the aim of investigating bulk dynamics free from wall-mode intrusions and accessing geostrophic regimes in which the intrinsic flow dynamics can develop unimpeded, while allowing different flow regimes, transitions, and large-scale structures to remain observable.
\\
\\
{\small
\noindent\textbf{Acknowledgements.}
The authors appreciate insightful discussions with Benjamin Favier, technical support from Freek van Uittert, Gerald Oerlemans, and Jørgen van der Veen, and experimental assistance from W. Vereijssen and R. Shetty during the early stages of this work.
\\
\\
\noindent\textbf{Funding.}
This work is part of the project “Universal critical transitions in constrained turbulent flows” (File No. VI.C.232.026) within the NWO-Vici research programme, funded by the Dutch Research Council (NWO). The authors are grateful for the support of NWO for the use of supercomputer facilities (Snellius) under grant no. 2025.011.
\\
\\
\noindent\textbf{Declaration of interests.}
The authors report no conflict of interest.
}

\begin{appen}
\section{DNS tests at more extreme parameter values}\label{tro}

As a test case, we performed DNS with parameters that matched those used in \cite{de2023robust}; see Fig. \ref{f0} and table \ref{t1}. These simulations mimic the extreme conditions encountered in the TROCONVEX facility  \citep{cheng2020laboratory,madonia2021velocimetry,madonia2023reynolds}, with low $Ek = 10^{-7}$ and a strong thermal forcing in the range $5.00 \times 10^{10} \leq Ra \leq 1.50 \times 10^{12}$ in a tall, slender cylinder with aspect ratio $\Gamma = 0.2$. Due to the substantial computational cost of these simulations, only periodic, cylindrical, and single-barrier configurations are considered. The ratio between the barrier width and the horizontal extent of the wall mode, $Ek^{1/3}$, was kept identical to that used in the fixed-$Ek$ cases presented in the main text (Fig. \ref{f2}(b), (c)). In this regime, wall modes have been shown to be particularly prominent; see \cite{de2023robust}. Consistent with this expectation and the observations discussed in the main text, although the barrier is able to disrupt the structure of the wall-mode, it does not fully suppress it and significant remnants are still present near the sidewall, as evidenced by vertical velocity fluctuations $Re_z$; see Fig. \ref{troc}(a). Similarly, the reduction in heat transport remains relatively modest, with the largest attenuation observed at the lowest $Ra$. However, as the difference between cylindrical and periodic cases increases with increasing $Ra$, the effectiveness of a single barrier becomes progressively reduced. This suggests that additional barriers may be required in these extreme regimes to further suppress sidewall-induced transport and narrow the gap between the cylindrical and periodic configurations. Experimental tests with more barriers and parameter scans are currently being planned in our laboratory.

\begin{figure}[ht!]
\centering
\includegraphics[width=1\textwidth]{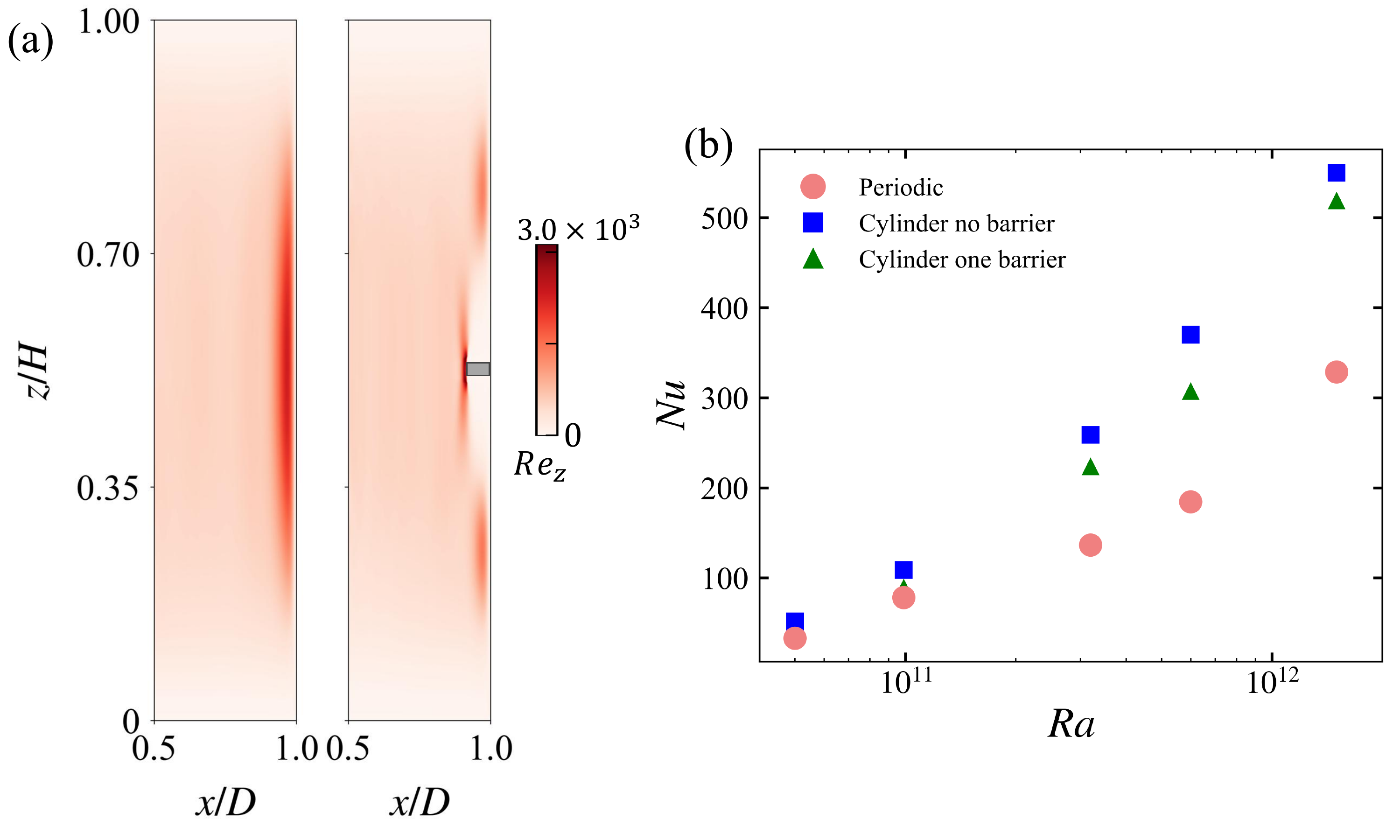}
\caption{\label{troc} (a) Azimuthally averaged fields of $Re_z$ for the no-barrier and one barrier configurations obtained from DNS at $Ra = 3.2 \times 10^{11}$ and $Ek = 10^{-7}$. The fields are shown in cylindrical coordinates, with the rotation axis located at $x/D = 0.5$ and the sidewall at $x/D = 1$. (b) Time-averaged $\langle Nu \rangle_t$ as a function of $Ra$ at fixed $Ek = 10^{-7}$.}
\end{figure}

\section{Baroclinic flow and effects of centrifugal buoyancy}\label{appA}

\begin{figure}[ht!]
\centering
\includegraphics[width=1\textwidth]{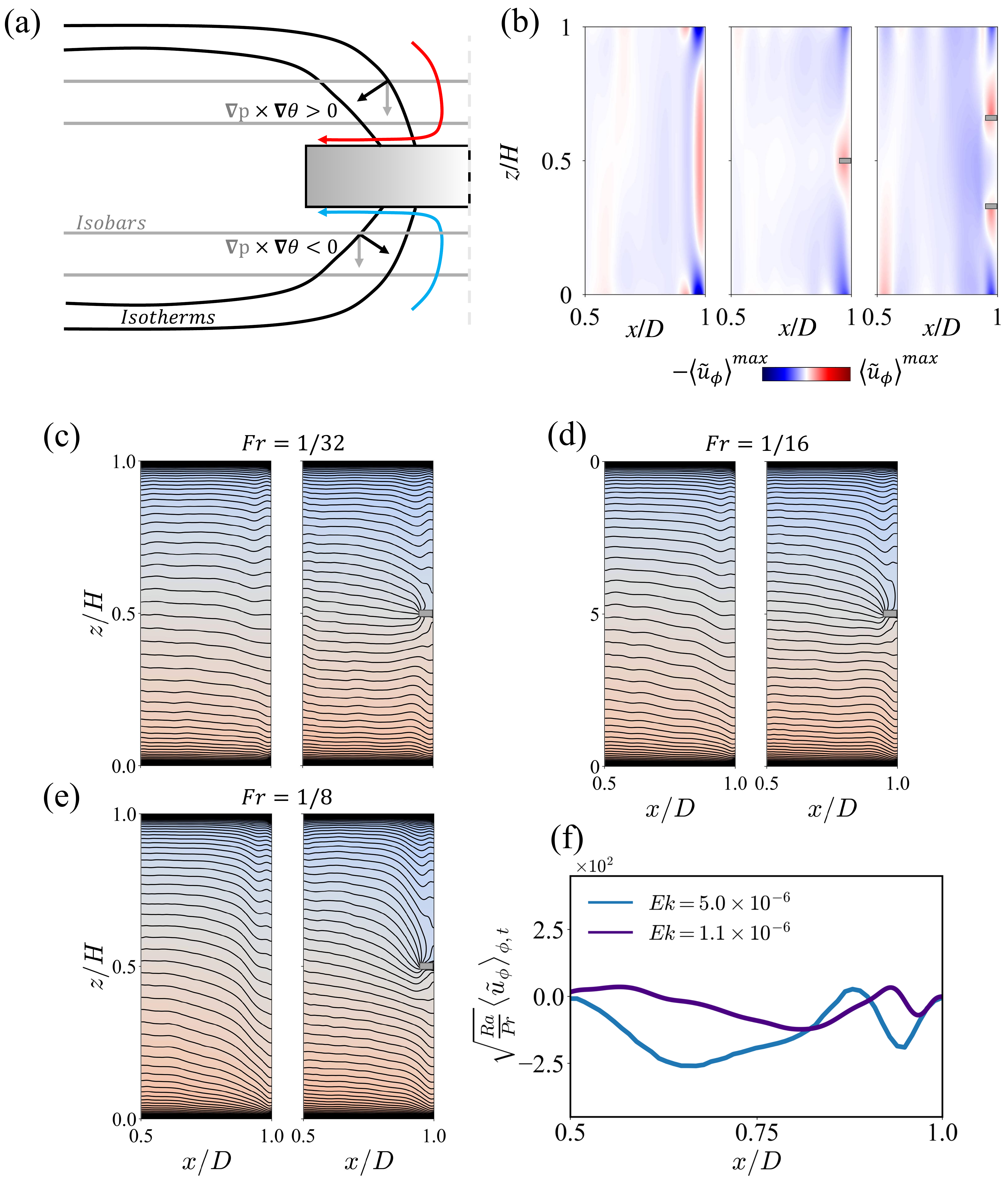}
\caption{\label{A} 
(a) Sketch showing isotherms and isobars near the barrier. The curvature of the isotherms, which sets the direction of the temperature gradient, determines the sign of the baroclinic vorticity above and below the barrier, owing to the corresponding sign of the cross product $\nabla p \times \nabla \theta > 0$ above the barrier and $\nabla p \times \nabla \theta < 0$ below.
(b) Heat maps of the time averaged azimuthal velocity from DNS for the no-barrier, one barrier, and two barrier cases at ${Ra=1.57\times10^9}$ and ${Ek=1.1\times10^{-6}}$. The fields are shown in cylindrical coordinates and are azimuthally averaged, with the symmetry axis at $x/D=0.5$ and the sidewall at $x/D=1$. (c)-(e) Temperature fields from DNS, overlaid with isotherms, for the no-barrier (left) and one barrier (right) configurations at different Froude numbers, indicating the relative magnitude of centrifugal forcing to gravity. (f) Time- and azimuthally averaged $\sqrt{Ra/Pr}~\tilde{u}_\phi$ profiles in regions located $5$ mm below the barrier for an experimental (blue curve) and a DNS case (purple curve), with $Fr = 1/16$. A negative azimuthal velocity is recovered below the barrier in the simulations, recovering agreement with the experiments. Note the difference in $Ek$ in bot cases.}
\end{figure}

In the main text, we discuss how the solid barrier locally distorts the isotherms, thereby inducing a steady baroclinic flow in its vicinity due to a mismatch with the pressure isobars (see Fig. \ref{f6}(d), (e)). We also note that in the experimental measurements, the flow above and below the barrier exhibits azimuthal directions opposite to each other (cyclonic and anticyclonic, respectively). This behaviour is not fully recovered in the DNS. Although a more detailed investigation of this flow would be valuable, it is beyond the scope of the present study; nevertheless, we provide some insights.

In figures \ref{f6}(d), (e), the isotherms bend symmetrically towards the barrier faces, whereas the isobars remain nearly unchanged. Calculations from DNS further reveal a misalignment between the pressure and thermal gradients, as illustrated schematically in figure \ref{A}(a). This mismatch gives rise to a source of baroclinic vorticity through the relation $D\boldsymbol\omega/Dt \sim \boldsymbol{\nabla} p \times \boldsymbol{\nabla} \theta$. The resulting mechanism produces positive vorticity above the barrier and negative vorticity below it. This asymmetric vorticity distribution drives a negative radial velocity along both face faces of the barrier (see the sketch in figure \ref{A}(a)), which is deflected by the Coriolis force in the counterclockwise direction, thereby explaining the symmetric azimuthal flow observed in the DNS, with positive azimuthal velocity on both sides of the barrier (Fig. \ref{A}(b)).

Figure \ref{A}(b) presents a global view of the azimuthal velocity field obtained from the DNS. The left panel corresponds to the case without barriers and shows the presence of a vertically coherent boundary zonal flow near the sidewall. After a single barrier (centre panel) is introduced, this vertically coherent structure is suppressed and is replaced by a positive azimuthal flow localised only in the vicinity of the barrier. This observation is also consistent with the experimental measurements away from the barrier vicinity, shown in Fig. \ref{f5}(c) of the main text. Finally, the right panel shows the two barriers configuration, where positive azimuthal velocities are again observed near both barriers, leaving a nearly neutral region between them, also in agreement with the experimental observations.

The reasons why this behaviour does not fully match the experimental observations are hypothesised to stem from low-order effects not included in the numerical simulations. In the DNS, the usual simplifying assumptions may limit the emergence of the observed asymmetry. To assess this, we include the typically neglected centrifugal buoyancy effect (figures \ref{A}(c)–(e)). Its magnitude relative to gravity is characterised by the Froude number, ${Fr} = \Omega^{2}R/g$. This contribution leads to a slight modification of the isotherms, as the cold fluid is driven outward, breaking the symmetry in the way the isotherms bend towards the barrier (figures \ref{A}(c–e)). Although small, this effect is already appreciable for the weak centrifugal forcing expected in the experiments, ${Fr} = 1/32$ (figure \ref{A}(d)). The asymmetry becomes progressively stronger as the centrifugal contribution increases. We identify this as the primary candidate symmetry-breaking mechanism. Centrifugal forcing preferentially enhances the positive vorticity source on the upper side of the system, leading to a stronger negative radial flow above the barrier and a correspondingly more pronounced positive azimuthal velocity in the upper region. In contrast, below the barrier the radial flow is weakened, resulting in near-zero or even positive radial velocities that dominate this region and give rise to the experimentally observed negative azimuthal flow below the barrier. This mechanism therefore provides a consistent explanation for the emergence of negative azimuthal velocity below the barrier and yields results consistent with the experiments, as shown in Fig. \ref{A}(f). We also note that additional mechanisms not included in the present simulations may further contribute to symmetry breaking, such as differences in thermal diffusivity between the barrier and the surrounding fluid.

More generally, achieving full quantitative agreement between DNS and experiments in relation to the baroclinic flows is challenging due to several small factors. A weak asymmetry is already visible in the DNS for the two barriers configuration (Fig. \ref{A}(b)), where a higher intensity is observed on the side closer to the top (bottom) plates, suggesting a sensitivity to small deviations from perfect symmetry. In experiments, slight misalignments, such as a small deviation of the barrier position from $z/H = 0.5$, may further enhance the observed asymmetry. Additional contributions may arise from imperfections in the barrier implementation. In practice, the barriers are not permanently attached to the cylindrical wall, but are carefully fitted to be flush with it, allowing them to be removed or repositioned if needed, which can leave small gaps that induce secondary flows. In DNS, the barriers are represented using the penalisation method explained in \S\ref{DNS}, which damps the velocities within the barrier region. Although this approach captures the main effect of a solid obstacle, more accurate representations may be required to fully mimic a real obstruction.

We conclude that weak effects, particularly the centrifugal buoyancy, typically treated as a low-order contribution and neglected, are likely responsible for the asymmetry observed in the experiments. Although weak, these effects appear to be essential for triggering the experimentally observed asymmetry regarding the baroclinic flow near the barrier.

\section{Tabulated parameters for simulations and experiments}\label{appB}
Table \ref{t1} summarises the parameters used in the DNS and the experiments.

\begin{table}
\centering

\begin{tabular}{cccccccc}
\toprule
\multicolumn{1}{c}{\textit{Simulations}} \\[3pt]
${Ra}$         & ${Ek}$         & ${Ro}$  &$\Gamma$       & $N_r \times N_\phi \times N_z$                   \\[3pt]
$1.57 \times 10^{9}$  & $2.52 \times 10^{-6}$ & $4.27 \times 10^{-2}$ &0.5& $193 \times 385 \times [577,769]$\footnotemark[2]  \\
$2.35 \times 10^{9}$  & $2.06 \times 10^{-6}$ & $4.27 \times 10^{-2}$ &                                                    \\
$3.51 \times 10^{9}$  & $1.69 \times 10^{-6}$ & $4.27 \times 10^{-2}$ &                                                    \\
$5.25 \times 10^{9}$  & $1.38 \times 10^{-6}$ & $4.27 \times 10^{-2}$ &                                                    \\
$7.85 \times 10^{9}$  & $1.13 \times 10^{-6}$ & $4.27 \times 10^{-2}$ &                                                    \\
$1.76 \times 10^{10}$ & $7.53 \times 10^{-7}$ & $4.27 \times 10^{-2}$ &                                                    \\
$3.93 \times 10^{10}$ & $5.04 \times 10^{-7}$ & $4.27 \times 10^{-2}$ &                                                    \\[3pt]
$1.57 \times 10^{9}$  & $1.13 \times 10^{-6}$ & $1.91 \times 10^{-2}$ &                                                    \\
$3.51 \times 10^{9}$  & $1.13 \times 10^{-6}$ & $2.85 \times 10^{-2}$ &                                                    \\
$7.85 \times 10^{9}$  & $1.13 \times 10^{-6}$ & $4.27 \times 10^{-2}$ &                                                    \\
$1.76 \times 10^{10}$ & $1.13 \times 10^{-6}$ & $6.39 \times 10^{-2}$ &                                                    \\
$3.93 \times 10^{10}$ & $1.13 \times 10^{-6}$ & $9.55 \times 10^{-2}$ &
                       \\
$5.88 \times 10^{10}$ & $1.13 \times 10^{-6}$ & $1.17 \times 10^{-1}$ &        
                    \\[3pt]
$5.00 \times 10^{10}$  & $1.00 \times 10^{-7}$ & $9.81 \times 10^{-3}$ &0.2& $385 \times 769 \times [961,1345]$\footnotemark[2]  \\
$9.90 \times 10^{10}$  & $1.00 \times 10^{-7}$  & $1.38 \times 10^{-2}$ &                                                    \\
$3.20 \times 10^{11}$  & $1.00 \times 10^{-7}$  & $2.48 \times 10^{-2}$ &                                                    \\
$6.00 \times 10^{11}$  & $1.00 \times 10^{-7}$  & $3.40 \times 10^{-2}$ &  &$577 \times 961 \times [1345,1537]$\footnotemark[2]                                                     \\
$1.50 \times 10^{12}$  & $1.00 \times 10^{-7}$  & $5.37 \times 10^{-2}$ &                                                    \\
                                 \\[3pt]

\multicolumn{1}{c}{\textit{Experiments}} \\[3pt]
${Ra}$         &  ${Ek}$        & ${Ro}$         &       $\Gamma$                                             \\[3pt]
$1.57 \times 10^{9}$  & $1.40 \times 10^{-6}$ &  $2.37 \times 10^{-2}$ &   0.5&                                                 \\
$3.06 \times 10^{9}$  & $1.40 \times 10^{-6}$ & $3.32 \times 10^{-2}$ &                                                    \\
$5.82 \times 10^{9}$  & $1.40 \times 10^{-6}$ & $4.63 \times 10^{-2}$ &                                                    \\
$8.28 \times 10^{9}$  & $1.40 \times 10^{-6}$ & $5.57 \times 10^{-2}$ &                                                    \\[3pt]
$1.57 \times 10^{9}$  & $1.40 \times 10^{-6}$ & $2.37 \times 10^{-2}$ &                                                    \\
$1.57 \times 10^{9}$  & $2.52 \times 10^{-6}$ & $4.27 \times 10^{-2}$ &                                                    \\
$1.57 \times 10^{9}$  & $5.10 \times 10^{-6}$ & $8.54 \times 10^{-2}$ &                                                    \\
\bottomrule
\end{tabular}
\caption{Input parameters for all experiments and simulations; grid resolution is specified for simulations only.}\label{t1}
\end{table}

\end{appen}
\clearpage

\bibliographystyle{jfm}
\bibliography{jfm}

\end{document}